\documentclass{aa}  
\usepackage{graphicx}
\usepackage{capt-of}
\usepackage{tabu}
\usepackage{txfonts}
\usepackage{ulem}
\usepackage{gensymb}
\usepackage{pifont}
\newcommand{\cmark}{\ding{51}}%
\newcommand{\xmark}{\ding{55}}%

\usepackage{amsmath}

\DeclareMathOperator{\atantwo}{atan2}
\usepackage{xcolor}

\newcommand{\xiv}{\boldsymbol{\xi}}
\newcommand{\cL}{\mathcal{L}}
\newcommand{\cB}{\mathcal{B}}
\newcommand{\cG}{\mathcal{G}}
\newcommand{\cR}{\mathcal{R}}

\newcommand{\rmd}{\mathrm{d}}
\newcommand{\Bv}{\mathbf{B}}
\newcommand\basisvec[1]{\hat{\boldsymbol{e}}_{#1}}
\newcommand{\bnabla}{\boldsymbol{\nabla}}
\newcommand{\solint}{\int_0^{R_{\odot}} \mathrm{d}r\,}
\newcommand{\wigred}[3]{\bigg(\begin{smallmatrix} \ell' & s & \ell \\ #1 & #2 & #3 \end{smallmatrix}\bigg)\,}

\newcommand{\tj}[6]{\left(\begin{array}{ccr}
#1 & #2 & #3\\
#4 & #5 & #6 \end{array}\right)}

\usepackage{geometry}
\usepackage{array}
\usepackage{tikz}

\usetikzlibrary{tikzmark,calc,positioning}

\newcommand{\DrawLine}[3][]{%
  \begin{tikzpicture}[overlay,remember picture]
    \draw[shorten <=-.2ex, shorten >=-.2ex,#1] (#2.north) -- (#3.south);
  \end{tikzpicture}
}

\newcolumntype{M}[1]{>{\centering\arraybackslash}m{#1}}

\renewcommand{\tikzmark}[2]{\tikz[overlay,remember picture,baseline] 
\node [anchor=base] (#1) {$#2$};}

\usepackage{hyperref}
\newcommand\orc[1]{\href{https://orcid.org/#1}{\includegraphics[width=3mm]{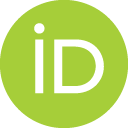}}}

\begin{document}

   \title{Unveiling complex magnetic field configurations in red giant stars}

   % \subtitle{I. Overviewing the $\kappa$-mechanism}

   % \author{G. Wuchterl
   %        \inst{1}
   %        \and
   %        C. Ptolemy\inst{2}\fnmsep\thanks{Just to show the usage
   %        of the elements in the author field}
   %        }

   % \institute{Institute for Astronomy (IfA), University of Vienna,
   %            T\"urkenschanzstrasse 17, A-1180 Vienna\\
   %            \email{wuchterl@amok.ast.univie.ac.at}
   %       \and
   %           University of Alexandria, Department of Geography, ...\\
   %           \email{c.ptolemy@hipparch.uheaven.space}
   %           \thanks{The university of heaven temporarily does not
   %                   accept e-mails}
   %           }

   % \date{Received September 15, 1996; accepted March 16, 1997}

      \author{S. B. Das\inst{1,2}\orc{0000-0003-0896-7972}
          \and
         {L. Einramhof}\inst{1}\orc{0009-0002-5619-0598}
          \and
         {L. Bugnet}\inst{1}\orc{0000-0003-0142-4000} 
          }

   \institute{
   Institute of Science and Technology Austria (ISTA), Am Campus 1, Klosterneuburg, Austria\\
           \email{srijanbdas@alumni.princeton.edu, lisa.bugnet@ist.ac.at} 
             \and Center for Astrophysics | Harvard \& Smithsonian,
              60 Garden Street, Cambridge, MA 02138, USA\\
}

   \date{Received XXXX; accepted ZZZZ}

% \abstract{}{}{}{}{} 
% 5 {} token are mandatory
 
  \abstract{{The recent measurement of magnetic field strength inside the radiative interior of red giant stars opens the way towards the full 3D characterization of the geometry of stable large-scale magnetic fields. However, current measurements, which are limited to dipolar $(\ell=1)$ mixed modes, do not properly constrain the topology of magnetic fields due to degeneracies on the observed magnetic field signature on such $\ell=1$ mode frequencies. Efforts focused towards unambiguous detections of magnetic field configurations are now key to better understand angular momentum transport in stars. We investigate the detectability of complex magnetic field topologies (as the ones observed at the surface of stars with a radiative envelope with spectropolarimetry) inside the radiative interior of red giants. We focus on a field composed of a combination of a dipole and a quadrupole (quadrudipole), and on an offset field. We explore the potential of probing such magnetic field topologies from a combined measurement of magnetic signatures on $\ell=1$ and quadrupolar ($\ell=2$) mixed mode oscillation frequencies. We first derive the asymptotic theoretical formalism for computing the asymmetric signature in {frequency pattern for $\ell=2$ modes} {due to} a quadrudipole {magnetic} field. To access asymmetry parameters for more complex magnetic field topologies, we numerically perform a grid search over the parameter space to map the degeneracy of the signatures of given topologies. We demonstrate the crucial role played by $\ell=2$ mixed modes in accessing internal magnetic fields with a quadrupolar component. The degeneracy of the quadrudipole {compared to pure} dipolar fields is lifted when considering {magnetic asymmetries in} both $\ell=1$ and $\ell=2$ {mode frequencies}. In addition to the analytical derivation for the quadrudipole, we present the prospect for complex magnetic field inversions using magnetic sensitivity kernels {from standard perturbation analysis for forward modeling}. Using this method, we explore the detectability of offset magnetic fields from $\ell=1$ and $\ell=2$ frequencies and demonstrate that offset fields may be mistaken for weak and centered magnetic fields, resulting in underestimating magnetic field strength in stellar cores. We emphasize the need to characterize $\ell=2$ mixed-mode frequencies, {(along with the currently characterized $\ell=1$ mixed modes)}, to unveil the higher-order components of the geometry of buried magnetic fields, and better constrain angular momentum transport inside stars.}}

   \keywords{stars: interior --- start: low mass --- stars:magnetic field --- stars: oscillations (including pulsations)}

   \maketitle
%
%--------t-----------------------------------------------------------

\section{Introduction}

The classic evolution picture of solar-like stars, with stellar cores spinning up after the end of core hydrogen burning on the main sequence, is today proven to be, for the most part, incorrect. Indeed, measurements from \cite{Beck2012, Mosser2012, Deheuvels2012a, Deheuvels2014a, Deheuvels2015, Deheuvels2017, Deheuvels2020,DiMauro2016a, Triana2017, Gehan2018, Tayar2019} indicate a relatively slow rotation rate in radiative interiors during advanced stages, incompatible with predicted rotation rates from classical hydrodynamic stellar models \citep[e.g.][]{Eggenberger2012, Ceillier2013, Marques2013a}. Among the key candidates to improve stellar evolution models and efficiently reproduce observations, are magnetic fields. When considering magnetohydrodynamic evolution involving the modified Tayler-Spruit dynamo formalisms \citep{Tayler1980, Spruit1999, Spruit2002, Mathis2005, Fuller2019, Eggenberger2022,Moyano2023, Petitdemange2023}, observed core and surface rotation rates can be reproduced simultaneously. In addition, stable fossil fields resulting from past convective dynamo action might be present inside the radiative core of red giants and may impact the angular momentum transport \citep[e.g.][]{Mestel1987, Duez2010}.
Theoretical predictions for the effect of magnetic fields in red giant stars' internal radiative zones on the frequencies of the oscillations have been developed during the last few years \citep{Loi2019, Loi2020, Loi2021, Gomes2020, Bugnet2021, Mathis2021, Li2022, Bugnet2022, Mathis2023}, and lead to the detection of several magnetized red giant cores by \cite{Li2022, Deheuvels2022, Li2023} and Hatt et al., \textsl{submitted}. These observed magnetic fields have, in common, a strong radial component up to a few hundred kilo Gauss in amplitude in the vicinity of the hydrogen-burning shell (H-shell). This is incompatible with current Tayler-Spruit formalisms generating strong toroidal components \citep{Fuller2019}. Observed magnetic fields must therefore have a different origin, and could result from the stabilization of past dynamo fields \citep[e.g.][]{Mestel1987, Braithwaite2008a, Duez2010, Bugnet2021}.

\begin{figure*}
    \centering
    \includegraphics[width=0.95\textwidth]{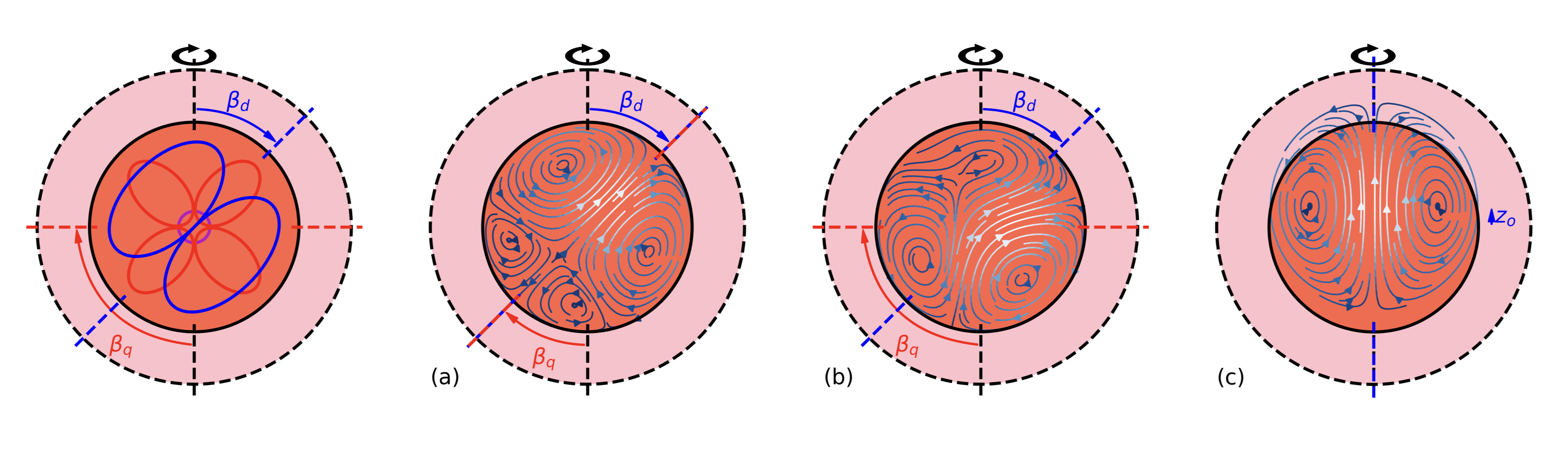}
    \caption{Left panel: Schematic diagram showing a dipole and quadrupole (quadrudipole) magnetic field where the dipolar field component (blue) is inclined at an angle $\beta_d$ from the rotation axis and the quadrupolar component (red) is inclined at an angle $\beta_q$ from the rotation axis. The purple circle indicates the hydrogen burning shell. The red area indicates the radiative interior and the pink area indicates the convective envelope (not to scale). The three rightmost panels represent the three magnetic field configurations used in our study as described in Section~\ref{sec:probe}. Case (a) shows a quadrudipole magnetic field with aligned dipolar and quadrupolar axes, inclined with the rotation axis of an angle $\beta=\beta_d=\beta_q$. Case (b) shows an inclined mixed quadrudipole with $\beta_d$ the angle between the rotation axis and the dipolar field, and $\beta_q$, the angle between the rotation axis and the quadrupolar field. Case (c) shows an offset dipole where the center of the dipole is shifted along the rotation axis by $z_o$ from the center of the star.}
    \label{fig:field_config_panel}
\end{figure*}

% {For instance, one scenario to explain the presence of magnetic white dwarfs is that magnetic fields might result from the conservation of fields trapped inside the radiative interior of red giant stars \citep[e.g.][]{Ferrario2020}. 

Magnetic fields at the surface of white dwarfs and intermediate-mass main-sequence stars are observed to be large-scale \citep[for instance in F stars][]{Seach2020, Zwintz2020}, with a dipolar poloidal field often dominating the spectropolarimetry results \citep[e.g.][]{Donati2009}. This geometry and associated strength are compatible with those of the radial magnetic field component detected in red giant cores \citep{Li2022, Deheuvels2022, Li2023}, and could therefore have resulted from the conservation of such a field in the radiative interior. However, not only dipolar but also quadrupolar and higher-order components are also very often detected \citep{Maxted2000, Euchner2005, Euchner2006, Beuermann2007, Landstreet2017}, and one pole can present stronger fields than the other \citep[hereafter called offset dipoles, e.g.][]{Wickramasinghe2000, Vennes2017, Hardy2023, Hardy2023a, Hollands2023}. This enhances the need for the characterization of the geometry of fields detected in red giants' internal radiative zones, as they might also not be pure dipoles. Accessing the magnetic field geometry inside the radiative interior during the red giant branch is key to understanding the origin of magnetic fields in white dwarfs, and to properly constrain the evolution of stars by including magnetic effects. Indeed, if the amplitude of the field controls how fast the angular momentum is redistributed, the geometry is key to constraining how much material is going to be redistributed in the radiative zone and therefore in the burning layers. 

%As the stellar plasma follows field lines \citep{Ferraro1937}, purely toroidal fields could, for instance, enhance less radial mixing than poloidal fields. High-order configurations (corresponding to small-scale fields) are likely to be less efficient than large-scale fields to redistribute the chemical composition from the radiative interior to the envelope. } %\textbf{Looks good.}

From the current measurements by \cite{Li2022, Deheuvels2022, Li2023}, we have access to the average radial magnetic field amplitude near the H-shell \citep[][]{Li2022, Bhattacharya2024}. Given the observed signatures in the frequency pattern, it is however currently impossible to confirm the complexity of the topology of the internal field (as it is done through spectropolarimetry for surface fields) for some of these stars, as the signature of the dipolar configuration on dipolar mixed modes is partially degenerate with higher order magnetic field configurations \citep[for instance with a field with a quadrupolar component, see][]{Mathis2023}.

We aim at lifting the observational degeneracy between magnetic field configurations from the use of combined constraints from dipolar ($\ell$=1) and quadrupolar ($\ell$=2) oscillations. In Section \ref{sec:methods} we present analytical and computational developments to link observed magnetic frequency asymmetries to the magnetic topology. In Section \ref{sec:probe} we investigate the detectability of various magnetic field configurations as observed at the surface of stars with a radiative envelope. %We study in particular the signature of a mixed dipolar and quadrupolar field and of an offset dipolar field shifted along the rotation axis. 
We discuss the detectability of mixed dipolar and quadrupolar configurations from a combined study of $\ell=1$ and $2$ oscillation frequencies, as well as the detectability study for offset magnetic fields along the rotation axis in Section~\ref{sec:discussion}. Finally, we conclude on the future potential of magnetoasteroseismology to unveil complex magnetic field topologies.

\section{Methods}
\label{sec:methods}

\subsection{Choice of the magnetic field configurations}

{Magnetic fields observed at the surface of stars with radiative envelopes often present a large-scale topology. While dipolar magnetic fields are observed \citep[e.g.][]{Donati2009}, higher-order and more complex configurations are also detected through spectropolarimetry. For instance, \cite{Kochukhov2022} observed a distorted dipolar topology at the surface of $\varphi$ Draconis, with a large inclination relative to the rotation axis and an asymmetry between the two magnetic poles. Cool Ap stars are known to exhibit even more complex magnetic field topology, such as 49 Cam that has significant octupolar contributions, including toroidal components \citep{Silvester2017}. We choose three magnetic field topologies, characterized by stable magnetic fields in the radiative zone, with low angular degree magnetic field configurations to be detectable using asteroseismic observables ($\ell=1$ and $\ell=2$ modes).} The three magnetic field configurations are represented in Fig.~\ref{fig:field_config_panel}, and discussed below. 

Case (a) is a magnetic field with dipolar component $\Vec{B}_d$ and quadrupolar component $\Vec{B}_q$ (hereafter quadrudipole, see second panel on Fig.~\ref{fig:field_config_panel}) having aligned magnetic axes ($\beta_d$ for the dipole is equal to $\beta_q$ for the quadrupole, as defined in the left panel of Fig.\ref{fig:field_config_panel}). For this case we allow the magnetic field axis to be inclined compared to the rotation axis with an angle $\beta=\beta_d=\beta_q$ (aligned dipole and quadrupole), and we write the ratio of the magnetic field strength of the quadrupole over the dipole as $\cR$. Hence,
\begin{equation}
\label{eq:aligned}
    \Vec{B}(r, \theta, \varphi) = \Vec{B}_d(r, \Tilde{\theta}, \Tilde{\varphi}) + \cR \, \Vec{B}_q(r, \Tilde{\theta}, \Tilde{\varphi}),
\end{equation}
% \sout{where the spherical grid $(\Tilde{\theta}, \Tilde{\varphi})$ in a coordinate system inclined at an angle $\beta$ from our reference coordinate where rotation is axisymmetric $(\theta, \Tilde{\varphi})$.} \textcolor{orange}
{where $(\theta, \varphi)$ are the spherical coordinates with respect to the rotation axis and $(\Tilde{\theta}, \Tilde{\varphi})$ the corresponding coordinates in a frame that is inclined at an angle $\beta$ compared to the rotation axis.} 

Case (b) is a misaligned quadrudipole (see third panel on Fig.~\ref{fig:field_config_panel}) where the dipolar and quadrupolar axes may have different inclination angles with respect to the rotation axis ($\beta_d \neq \beta_q$):
\begin{equation}
\label{eq:misaligned}
    \Vec{B}(r, \theta, \varphi) = \Vec{B}_d(r, \Tilde{\theta}_d, \Tilde{\varphi}_d) + \cR \, \Vec{B}_q(r, \Tilde{\theta}_q, \Tilde{\varphi}_q) \, ,
\end{equation}
where, the coordinate system $(\Tilde{\theta}_d, \Tilde{\varphi}_d)$ is inclined by $\beta_d$ with respect to the rotation axis and the coordinate system $(\Tilde{\theta}_q, \Tilde{\varphi}_q)$ is inclined by $\beta_q$ with respect to the rotation axis. A recipe for constructing the rotated dipole and quadrupole is laid out in Appendix~\ref{sec:Br_dip_quad} for the readers' convenience. We use simple identities offered by Wigner $d$-matrices for the necessary transformation between spherical coordinate systems.

Case (c) is an offset dipolar field, with an offset along the rotation axis (see case (c) in Fig. \ref{fig:field_config_panel}), resulting in one pole more magnetized than the other as observed at the surface of white dwarfs \citep[e.g.][]{Wickramasinghe2000}:

\begin{equation}
\label{eq:offcentered}
    \Vec{B}(r, \theta, \varphi) = \Vec{B}_d(\tilde{r}(z_o), \tilde{\theta}(z_o), \varphi),
\end{equation}
where
\begin{gather}
    \tilde{r}(z_o) = \sqrt{r^2 + z_o^2 -2r\cos{\theta}z_o}, \\
    \tilde{\theta}(z_o) = \atantwo(r\sin{\theta}, r\cos{\theta}-z_o) \, ,
\end{gather}
and $z_o$ the offset of the center of the field along the polar axis.

\subsection{General definition of asymmetry parameters $a_{\ell |m|}$}

To investigate the detectability of these various large-scale magnetic field topologies inside the radiative interior of red giants, we use the asymmetry they induce on $\ell=1$ and $\ell=2$ mixed-mode frequencies \citep[as demonstrated in][]{Bugnet2021, Li2022}. We consider a slowly rotating star, to ensure that the rotation can be treated as a first-order perturbation (valid for slow rotators like red giants) and weakly magnetized following the derivation in \cite{Bugnet2021} (the magnetic field can as well be treated as a first-order perturbation). On carrying out a linearization of the magnetohydrodynamic equations about a magnetostatic background state, it can be shown that the Lorentz-state operator $\cL$ affecting oscillation frequencies may be expressed as \citep[see Appendix~A in][for further details]{Das2020}:
\begin{eqnarray}
    4\pi\,\delta \cL \xiv =  \mathbf{B}_0 \times (\bnabla \times \mathbf{B}_1) - (\bnabla \times \mathbf{B}_0) \times \mathbf{B}_1 - \bnabla [\xiv \cdot (\mathbf{j}_0 \times \mathbf{B}_0)] \, ,
\end{eqnarray}
where $\Bv_0$ and $\Bv_1$ are the background and perturbed magnetic field respectively, $\xiv$ the eigenstate of the mode, and $\mathbf{j}_0 = \bnabla \times \Bv_0$ is the background current density.

For a high radial order $g$-dominated mixed mode, the coupling of eigenstate $\xiv_{\ell, m}$  of a mode labelled by angular degree $\ell$ and azimuthal order $m$ with eigenstate $\xiv_{\ell,m'}$ due to the magnetic linear operator $\cL$ is given by a magnetic coupling matrix $\mathbf{M}$ \citep{Das2020, Li2022}, such as
\begin{equation} \label{eq:eigenvalue_problem}
     %\mathbf{M}_{\ell} \, \mathbf{v} = \mathbf{\mathcal{W}}^1 \,  \mathbf{v} \, ,
     \mathbf{M}_{\ell} \mathbf{v}_{\ell, k} = \omega^{1}_{\ell, k} \mathbf{v}_{\ell, k}
\end{equation}
where $\omega^1_{\ell, k}$ are the $2\ell+1$ linearly perturbed eigenfrequencies corresponding to mode $\ell$, $\mathbf{v}_{\ell, k}$ are the corresponding eigenvectors, and the elements of the mode coupling matrix $\mathbf{M}_{\ell}$ are defined as
%where $\mathbf{\mathcal{W}}^1$ is a diagonal matrix containing the linearly perturbed eigenfrequencies $\omega^1_{\ell, m}$ for the mode $(\ell, m)$, $\mathbf{v}$ are the corresponding eigenvectors and the elements of the mode coupling matrix $\mathbf{M}_{\ell}$ are defined as
\begin{equation}
    M_{\ell}^{m,m'}=
    \frac{\left<{\boldsymbol\xi}_{\ell,m},{\cL}\left({\boldsymbol\xi}_{\ell,m'}\right)\right>}{2\omega^0_{\ell}\left<{\boldsymbol\xi}_{\ell,m'},{\boldsymbol\xi}_{\ell,m}\right>}\, .
\label{eq:MatrixLorentz}
\end{equation}
\noindent where $\omega^0_{\ell}$ is the unperturbed frequency of the mixed mode of order $\ell$. %\sout{All three quantities $\mathbf{M}, \mathbf{v}_k,$ and $\omega^1_k$ in Eq.~\ref{eq:eigenvalue_problem} depend on $\ell$, which we chose to suppress in the notation.}\textcolor{red}{we need the l dependency in the appendix} In general, each $k$ is a mixture of all azimuthal orders $m$ and cannot be uniquely identified with a single $m$.
As expressed in \cite{Li2022}, and further supported by \cite{Bugnet2021} in the axisymmetric case, the magnetic field’s presence induces asymmetries in mixed mode multiplets. In our study, we consider that magnetic field effects are smaller than rotational effects, resulting in a ${\bf M}_{\ell}$ matrix with dominant diagonal terms \citep[we refer to][for the derivation of the full coupling matrix which includes effects of inclination of magnetic axis with respect to rotation axis]{Loi2021}. This is a reasonable assumption, as observed magnetic fields by \cite{Li2022, Deheuvels2022, Li2023} from $\ell=1$ frequencies are detected on multiplets containing $2\ell+1$ components, and not on $(2\ell+1)^2$. Thus, by neglecting the off-diagonal terms, $\mathbf{M}_{\ell}$ becomes a diagonal matrix, with the eigenfrequencies on the diagonal. In this case, $k=m$ and we write ($\mathbf{v}_{\ell,m}, \omega^1_{\ell,m}$) instead. %Loosing the perturbation exponent for convenience ($\omega^1_{\ell,m}=\omega_{\ell,m}$), 
We generalize the formalism of \cite{Li2022} for the asymmetry induced by magnetic fields on $\ell=1$ mixed mode frequencies for any $\ell$ modes as:
\begin{equation} \label{eqn: asym_param_def}
    \delta_{{\ell, m}_{\rm asym} }= \omega_{\ell, -m} + \omega_{\ell, m} - 2 \omega_{\ell, 0} = \left({2}\ell+1\right) \zeta a_{\ell |m|} \omega_{B}^{\ell}\, .
\end{equation}

\noindent In Eq.~\ref{eqn: asym_param_def}, $\zeta$ is the coupling function of the mixed modes \citep{Goupil2013} and $a_{\ell |m|}$ the asymmetry parameter is defined as:%$m \in [-\ell, \ell]$, and
\begin{equation} \label{eqn: asymmetry_parameter}
    a_{\ell |m|} = \frac{M_{\ell}^{|m|,|m|} + M_{\ell}^{-|m|,-|m|} - 2 M_{\ell}^{0,0}}{\mathrm{Tr}\left(\mathbf{M}_{\ell}\right)} \, 
\end{equation}
\noindent \noindent with  ${\mathrm{Tr}\left(\mathbf{M}_{\ell}\right)}=\sum_{m=-\ell}^\ell M_{\ell}^{m,m}$, and $\omega_{B}^{\ell}$ the mean frequency shift:
\begin{equation}
    \omega_{B}^{\ell} = \frac{{\mathrm{Tr}\left(\mathbf{M}_{\ell}\right)}}{2\ell+1}\, .
\end{equation}

\begin{figure}
\centering
    \includegraphics[width=1\linewidth]{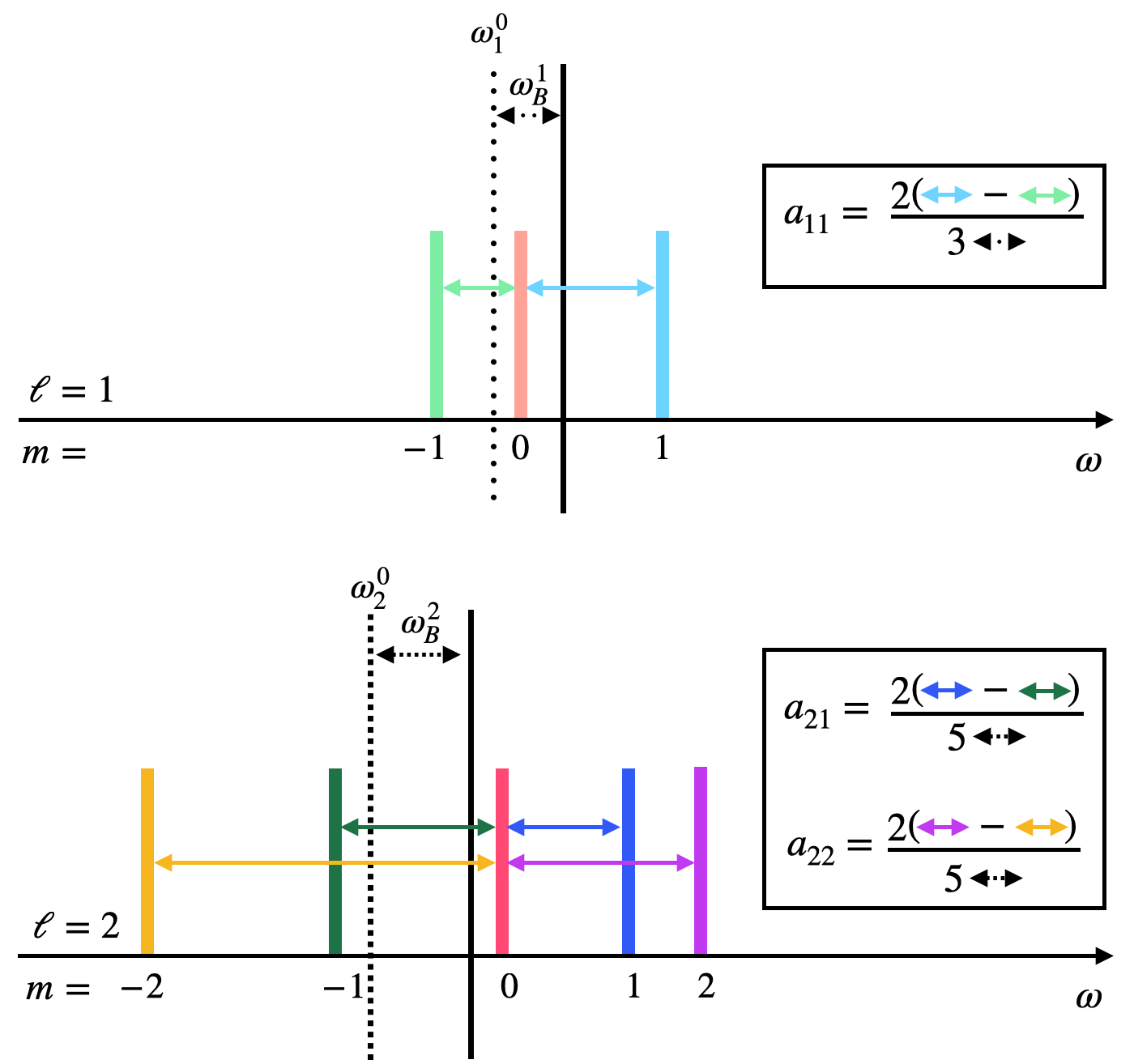}
    \caption{Definition of the asymmetry parameters $a_{\ell|m|}$ for $\ell=1$ \citep[top, as in][]{Li2022} and $\ell=2$ (bottom) oscillation multiplets.}
    \label{fig:scheme}
\end{figure}

In Fig.~\ref{fig:scheme} are represented the definitions of the three asymmetry parameters used in our study, for $\ell=1$ ($a_{11}$) and $\ell=2$ ($a_{21}$, $a_{22}$) oscillations. In the following subsections we outline the two methods we have used to calculate magnetic frequency splittings (and hence asymmetry parameters) --- (i) the analytical approach, similar to \cite{Mathis2023}, which adheres to the simplifying assumptions mentioned in the supplementary section S2.2 of \cite{Li2022} and is sensitive to only the $(\theta, \varphi)$ angular dependence of $B_r^2$ and (ii) the numerical approach using \texttt{magsplitpy} \citep[a rigorous computational framework for computing magnetic splittings due to a general magnetic field, following the theoretical underpinnings of][]{Das2020}, which is sensitive to all components of magnetic fields and provides the full solution.

\subsection{Analytical approach: probing $B_r$}
For a magnetic field $\Bv = (B_r, B_{\theta}, B_{\varphi})$, it is known that the dominant contribution to observed %$\ell=1$ 
g-dominated mixed-mode frequency splitting comes from the $B_r^2$ component \citep[][]{Bugnet2021, Mathis2021} in the vicinity of the H-shell \citep[][]{Li2022, Bhattacharya2024}. As shown in Eq.~30 of \cite{Li2022}, the elements of this coupling matrix when considering only the dominant magnetic term can be approximated as: 
%under these simplifying assumptions can be expressed as  
\begin{align} \label{eqn: M_analytical}
    M_{\ell}^{m,m'} &= \frac{1}{\mu_0} \int_{r_i}^{r_o} \left[\frac{\partial(r \xi_h)}{\partial r}\right]^2 \int_{0}^{2\pi} \int_0^{\pi} B_r^2 e^{i(m'-m)\varphi} \nonumber \\
    & \times \quad \left[ \frac{\partial \hat{Y}_{\ell m}}{\partial \theta} \, \frac{\partial \hat{Y}_{\ell m'}}{\partial \theta} + \frac{mm'}{\sin^2{\theta}} \hat{Y}_{\ell m} \, \hat{Y}_{\ell m'}\right] \sin{\theta} \, \rmd r \, \rmd \theta \, \rmd \varphi \, ,
\end{align}
where $\xi_h$ is the radial variation of the horizontal component of the eigenfunction,  $Y_{\ell m}(\theta, \varphi) = \hat{Y}_{\ell m}(\theta) \, e^{i m \varphi}$ and $r_i, r_o,$ are the inner and outer turning points of the $g$-mode cavity. Using this analytical expression and the Racah-Wigner algebra derived in Appendix B of \cite{Mathis2023}, we obtain expressions for the various $M_{\ell}^{m,m'}$ (see \cite{Mathis2023} for $\ell=1$ modes, and our Appendix~\ref{app:matrix} for $\ell=2$ modes). From this, we obtain analytical expressions of the asymmetry parameters for various radial magnetic field topologies in Section \ref{sec:probe}. 
 
% \textcolor{blue}{add a sentence saying that the theoretical work comes in expansion/analogy of Mathis \& Bugnet 2023}

\subsubsection{A simple radial quadrudipole for theoretical calculations}
\label{sec:quadrudipole_th}

Appendix~A outlines the steps to obtain a rotation of a field on the surface of a sphere --- going from being axisymmetric on the original coordinate system $(\theta, \varphi)$ to being non-axisymmetric on $(\theta, \varphi)$. As shown, we use identities of Wigner-$d$ matrices to do so. In our case, we use the same analytical steps to tilt the axis of a dipolar and quadrupolar field by $\beta_d$ and $\beta_q$ respectively to obtain the expression for a general quadrudipole where the dipole and quadrupole are not necessarily axis-aligned. Simplifying the Wigner-$d$ matrices required to tilt the field, we obtain the following analytical expression,  expanding on the coaligned case from \cite{Mathis2023}:
\begin{align} \label{eqn: quadrudipole_Br_expanded}
    B_r(r,\theta,\varphi, &\mathcal{R}, \beta_d, \beta_q) = B_0 \, b_r(r) \, \frac{1}{2}\sqrt{\frac{3}{\pi}}  \nonumber \\ & \Big[\cos{\beta_d}\,\cos{\theta}
    +  \sin{\beta_d} \, \sin{\theta} \, \cos{\varphi}\\\nonumber 
    &+ \frac{\sqrt{15}}{4} \mathcal{R} \Big(\frac{1}{3} -  \cos^2{\beta_q} - \cos^2{\theta}
    + 3 \cos^2{\beta_q} \, \cos^2{\theta}\\\nonumber 
    &+ \sin{2\beta_q} \, \sin{2\theta} \, \cos{\varphi} + \sin^2{\beta_q} \, \sin^2{\theta} \, \cos{2\varphi}\Big)\Big] \, ,
\end{align}
%\textcolor{blue}{Can we replace this expression with the more general one derived by Lukas, adding the shift from the rotation axis? This way, we can then develop into co-aligned and centered fields, non-aligned and centered, and off-centered,... see subsections below.}
where $\mathcal{R}$ is the ratio of the strength of quadrupole to dipole as defined in \cite{Mathis2023}. {Here, we explicitly assume that both the dipole and the quadrupole have the same radial dependence $b_r(r)$. This is a reasonable assumption since the dominating contribution from the magnetic field on the mode splitting occurs in the vicinity of the H-shell \citep[][]{Li2022, Bhattacharya2024}, hence the radial profile of the field does not significantly affect the magnetic field signature.} We use this formalism in Section~\ref{sec:probe} to compute asymmetry parameters related to $B_r^2$ associated with various magnetic field topologies trapped in the radiative interior of the star. %, $\beta_d, \beta_q$ are the angles of inclination of the dipole and quadrupole to the rotation axis respectively. 
For an aligned quadrudipole (Case (a), such that $\beta = \beta_d = \beta_q$) this reduces to Eq.~25 and Eq.~28 of \cite{Mathis2023} for $\mathcal{R}=0$ and $\beta=0$, respectively. 
% \sout{In Section~\ref{sec:probe}, we use this simplified radial magnetic field expression to theoretically evaluate the detectability of the radial component of the magnetic field for Cases (a) \& (b)}.

\subsection{\texttt{magsplitpy}: implementation of the full system} \label{sec: magsplitpy}

To estimate the signature of magnetic field topologies which are more complex than a quadrudipole, deriving an analytical expression is no longer efficient. In the same spirit as the inversion of rotation rates inside red giant stars from mixed-mode splitting \citep[e.g.][]{Deheuvels2012a, DiMauro2016a, Ahlborn2020, Pijpers2021}, \cite{Das2020} propose sensitivity kernels to probe a general magnetic field topology. 

\cite{Das2020} formulated a prescription to infer the global solar magnetic field by using tools prevalent in terrestrial seismology \citep{Dahlen1999}. The \cite{Das2020} formalism is general enough to be seamlessly applied to other stars whose internal structure (and hence mode eigenfunctions) can be calculated from stellar evolution codes.
%such as \texttt{MESA} \citep{Paxton2013, Paxton2019} and \texttt{GYRE} \citep{Townsend2014a}. 
%This appendix \textcolor{blue}{which one?} outlines the salient points of this method for ease of reference of the reader.
% The numerical code \texttt{magsplitpy} accepts stellar model as input (in the form of mode eigenfunctions and eigenfrequencies) and provides us all the kernel components for a general field topology (see Appendix C in \cite{Das2020} for an extensive list of kernel expressions). We also refer to \textcolor{blue}{Shatanik paper} for an extensive probe of magnetic field sensitivity kernels in red giant stars. 
In our study, we model a typical red giant star using \texttt{MESA} \citep{Paxton2011, Paxton2013, Paxton2015, Paxton2018, Paxton2019, Jermyn2023} and compute its eigenfunctions and eigenfrequencies using \texttt{GYRE} \citep{Townsend2013, Townsend2014a}. The \texttt{MESA}\footnote{The corresponding \texttt{MESA inlist} file is available on Zenodo at \textit{The folder will be uploaded at the time of publication}} computation is initialized with a mass of $1.5M_\odot$ and metallicity of $Z=0.02$. We extract the model for which $\Delta\nu = 14.49\mu$Hz, which represents a typical red giant branch star. We define $R_h$ as the radius where the pp-nuclear reaction reaches its maximum, while $R_c$ is the radius at which the Brunt-Väisälä frequency first goes to zero.

\subsubsection{Magnetic inversion kernels}
 
  % In the same spirit as the inversion of rotation rates inside red giant stars from mixed-mode splitting \citep[e.g.][]{Deheuvels2012a, DiMauro2016b, Ahlborn2020, Pijpers2021}, \cite{Das2020} propose sensitivity kernels to probe a general magnetic field topology. 
  %Given a general 3D magnetic field $\Vec{B}$, D20 calculates sensitivity kernels for computing frequency splittings corresponding to this field. 
  Since the Lorentz force is given by $(\boldsymbol{\nabla} \times \mathbf{B}) \times \mathbf{B}$, the perturbation of interest are the components of the second rank Lorentz-stress tensor $\boldsymbol{\mathcal{H}} = \mathbf{BB}$. %\textcolor{blue}{Please explain this Srijan, the community is not used to the Lorentz stress tensor.} 
  Therefore, to decompose these tensors in a spherical geometry, \cite{Das2020} uses generalized spherical harmonics \citep[GSH as in Appendix~C of][]{Dahlen1999} $Y_{st}^{\mu}(\theta,\varphi)$ such as
 \begin{eqnarray}
     \mathbf{B}(r,\theta,\varphi) &=& \sum_{s=0}^{\infty} \sum_{t=-s}^{s}\sum_{\mu} B_{st}^{\mu}(r)\, Y_{st}^{\mu}(\theta,\varphi) \,\basisvec{\mu} \label{eqn: B_exp_GSH},\\
     \boldsymbol{\mathcal{H}}(r,\theta,\varphi) &=& \sum_{s=0}^{\infty} \sum_{t=-s}^{s} \sum_{\mu\nu} h_{st}^{\mu \nu}(r)\, Y_{st}^{\mu + \nu}(\theta,\varphi)\, \basisvec{\mu}\, \basisvec{\nu} \, .\label{eqn: H_exp}
 \end{eqnarray}
where $(\mu, \nu) \in \{-1, 0, +1\}^2$ and $s, t$ subscripts denote the spherical harmonic angular degree and azimuthal order. Note that in this study, we use $\ell, m$ for mode harmonics and $s, t$ for perturbation harmonics. The basis vectors in spherical polar coordinates can be transformed to those in the GSH basis using the following transformation
 \begin{equation}
     \basisvec{-} = \frac{1}{\sqrt{2}}(\basisvec{\theta} - i \basisvec{\varphi}), \qquad \basisvec{0} = \basisvec{r}, \qquad \basisvec{+} = -\frac{1}{\sqrt{2}}(\basisvec{\theta} + i \basisvec{\varphi}).
\end{equation}
for brevity of subscripts \citep[and in keeping with the convention of][]{DT98}, we denote $\mu = -1, +1$ in the subscripts as $\basisvec{-}, \basisvec{+}$ respectively.

As a result, the general elements of the matrix $\mathbf{M}_{\ell}$ write:
\begin{align}\label{eqn:lamda_decomp}
    M_{\ell}^{m,m'} &= \sum_{st} \sum_{\mu\nu} \solint r^2{}_{mm'}\cB_{st}^{\mu\nu} (r) \,h_{st}^{\mu\nu} (r)
\end{align}
with $h_{st}^{\mu\nu}$ are the Lorentz-stress tensors for components $(\mu,\nu)$ and spherical harmonic $(s,t)$ while ${}_{mm'}\cB_{st}^{\mu\nu}$ are the respective magnetic inversion kernels. The complete expressions of the kernel components ${}_{mm'}\cB_{st}^{\mu\nu}$ is laid out in Appendix~\ref{sec: Brsq_kerns} \citep[presented for ease of readers' reference but originally found in][]{Das2020}. A similar expression was derived in \cite{Mathis2023}. Note that since we are confined to the self-coupling of multiplets (same $n, \ell$ coupling), we suppressed these indices in the above expression.

\subsubsection{Sensitivity of modes with degree $\ell$ to the magnetic field topology}
\label{sec:sensitivity_ell}

To ask the question of which components of the magnetic field $\Bv$ are sensitive modes of degree $\ell$, we need to see how the Lorentz-stress GSH is connected to the magnetic field GSH. This is because modes are sensitive to components of $\boldsymbol{\mathcal{H}}$. As shown in Appendix~D of \cite{Das2020}, they are related as follows:
\begin{eqnarray} 
    h^{\mu \nu}_{s t} &=& \sum_{s_1,s_2,t_1,t_2} B_{s_1 t_1}^{\mu} B_{s_2 t_2}^{\nu} \int Y^{* \mu + \nu}_{s t} Y_{s_1 t_1}^{\mu} Y_{s_2 t_2}^{\nu} \mathrm{d}\Omega \nonumber \\
    &=& \sum_{s_1,s_2,t_1,t_2}B^{\mu}_{s_1 t_1}B^{\nu}_{s_2 t_2} (-1)^{\mu + \nu + t} \sqrt{\frac{(2s+1)(2s_1 + 1)(2s_2 + 1)}{4 \pi}} \nonumber \\
    &\times&\tj{s_1}{s}{s_2}{\mu}{-(\mu+\nu)}{\nu} \tj{s_1}{s}{s_2}{t_1}{-t}{t_2} \, . \label{eqn: generic_Hmunu}
\end{eqnarray}

\noindent Whether or not a degree of perturbation $s$ will induce a frequency splitting, depends on the angular degree $\ell$ of the mode of interest. This is controlled by the triangle rule imposed by the Wigner 3-$j$ symbols in Eq.~\ref{eqn: generic_Hmunu}. 
For simplicity, we assume that the magnetic field is a pure dipole, i.e., $s_1=s_2=1$. By the Wigner 3-$j$ triangle rule $|s_1 - s_2| \leq s \leq s_1 + s_2$, there would be only three degrees of Lorentz-stress tensor $s=0, 1, 2$. When using self-coupling of $\ell=1$ modes, the odd degree $s=1$ is insensitive since the kernel ${}_{\ell \ell} \cG_{1}^{00} = 0$ (note that $\cG$ are the $m$ independent forms of the full kernels $\cB$, see Appendix~\ref{sec: Brsq_kerns}). So, for self-coupling, $\ell=1$ modes are only sensitive to $s=0,2$ components of Lorentz-stress. The magnetic field component $s_1=1$ contributes to both these Lorentz-stress components. Therefore, when using dipole modes $\ell=1$ (resp. quadrupole modes $\ell=2$), the splittings and asymmetry parameters are sensitive
to only up to $s=2$ (resp. $s=4$) components of the Lorentz-stress tensor.

Therefore, the important conclusion from the above thought experiment is that we can only infer even components of the Lorentz stress. However, each of these components, let's say $s=2$, contains contributions from all magnetic field components $s/2~\leq~s_1~\leq~\infty$. So, the $s=0$ Lorentz-stress has information from all $B_r$ components with angular degree $s_1 \geq 0$ (in reality $s_1 \geq 1$, since $s = 0$ is a magnetic monopole). Similarly, the $s=2$ Lorentz stress component has information from all $B_r$ components with angular degree $s_1 \geq 1$, the $s=4$ Lorentz stress component has information from all $B_r$ components with angular degree $s_1 \geq 2$, and so on and so forth. Quadrupolar modes ($\ell=2$) are therefore extremely valuable for the search of complex magnetic field topologies, as $\ell=2$ oscillation mode frequencies are independent of the dipolar component of the magnetic field, and give a direct insight into the high-order of complexity of the field. This is the foundation of this study and justifies the need for $\ell=2$ mode characterization in the following sections.\\

\subsubsection{Realistic quadrudipole configurations for signatures of the full magnetic field with \texttt{magsplitpy}}
\label{sec:quadrudipole_full}

In order not only to check the theoretical results obtained from the simplified radial component of a quadrudipole but also to investigate the signature of more complex topologies in Section~\ref{sec:probe}, we use a full quadrudipole topology in \texttt{magsplitpy}. Deriving a force-free stable quadrudipole magnetic field in the radiative interior from \cite{Broderick2007}:

\begin{align}\label{eq:dpol_full}
    %\Vec{B}^{r<R_c}_d(r, \theta, \varphi)
    \Vec{B}_d(r, \theta, \varphi)&= C_d \bigg[\frac{j_1(\alpha_d r)}{r} \cos\theta \, \basisvec{r} \nonumber\\
    &\qquad - \frac{\alpha_d r j_0(\alpha_d r) - j_1(\alpha_d r)}{2r} \sin\theta \, \basisvec{\theta} \nonumber\\
    &\qquad + \frac{\alpha_d j_1(\alpha_d r)}{2}\sin\theta \, \basisvec{\varphi}\bigg], 
\end{align}
and
\begin{align}\label{eq:qpole_full}
    % \Vec{B}^{r<R_c}_q(r, \theta, \varphi) 
    \Vec{B}_q(r, \theta, \varphi) &= C_q \bigg[\frac{j_2(\alpha_q r)}{6r} (3\cos^2\theta - 1)\, \basisvec{r} \nonumber\\
    &\qquad - \frac{\alpha_q r j_1(\alpha_q r) - 2 j_2(\alpha_q r)}{6r} \cos\theta\sin\theta \, \basisvec{\theta} \nonumber\\
    &\qquad + \frac{\alpha_q j_2(\alpha_q r)}{6}\cos\theta\sin\theta \, \basisvec{\varphi}\bigg],
\end{align}

\noindent where $j_{l\in [1,2]}$ are the spherical bessel function of the first kind. The parameters $\alpha_d$ and $\alpha_q$ are chosen such that the conditions at the convective/radiative boundary are $B_r(R_c) = B_\varphi(R_c) = 0$ with $R_c$ the radius of the radiative interior \citep[see][for more details about the method]{Prat2019, Bugnet2021}. As a result, both the quadrupolar and the dipolar field are zero outside the radiative region, i.e., $\Vec{B}({r\geq R_c}) = \Vec{0}$. Both fields are normalized such that $B_r(R_h) = 1$, where $R_h$ is the radius of the H-shell in the radiative zone. Therefore, for the full numerical calculations using \texttt{magsplitpy}, we use the following 3D field:
\begin{equation}
    \Vec{B}(r, \theta, \varphi) = \Vec{B}_d(r, \theta, \varphi) + \mathcal{R} \, \Vec{B}_q(r, \theta, \varphi) \, .
\end{equation}
In this general field configuration, we define $\cR$ as the relative strength between radial components of the dipole and quadrupole at the H-shell. This is a reasonable choice because as evaluated in \cite{Bhattacharya2024} \citep[and supported by][]{Li2022}, the hydrogen burning region represents about 90\% of the sensitivity of the $\ell=1,2$ oscillation modes in the code for the typical model red giant star chosen in our study and the radial component dominates the magnetic sensitivity over all other components. This quadrudipole general formalism is used in Section~\ref{sec:probe} to estimate the detectability of Cases (a), (b), and (c) from $\ell=1$ and $\ell=2$ oscillation asymmetries with \texttt{magsplitpy}.

\section{Asymmetry parameters as a probe of magnetic field topologies}\label{sec:probe}

\subsection{{Aligned dipolar field and quadrupolar field axes}}\label{sec:aligned_quadrudipole}

\begin{figure*}[ht]
    \centering
    \includegraphics[width=0.94\textwidth]{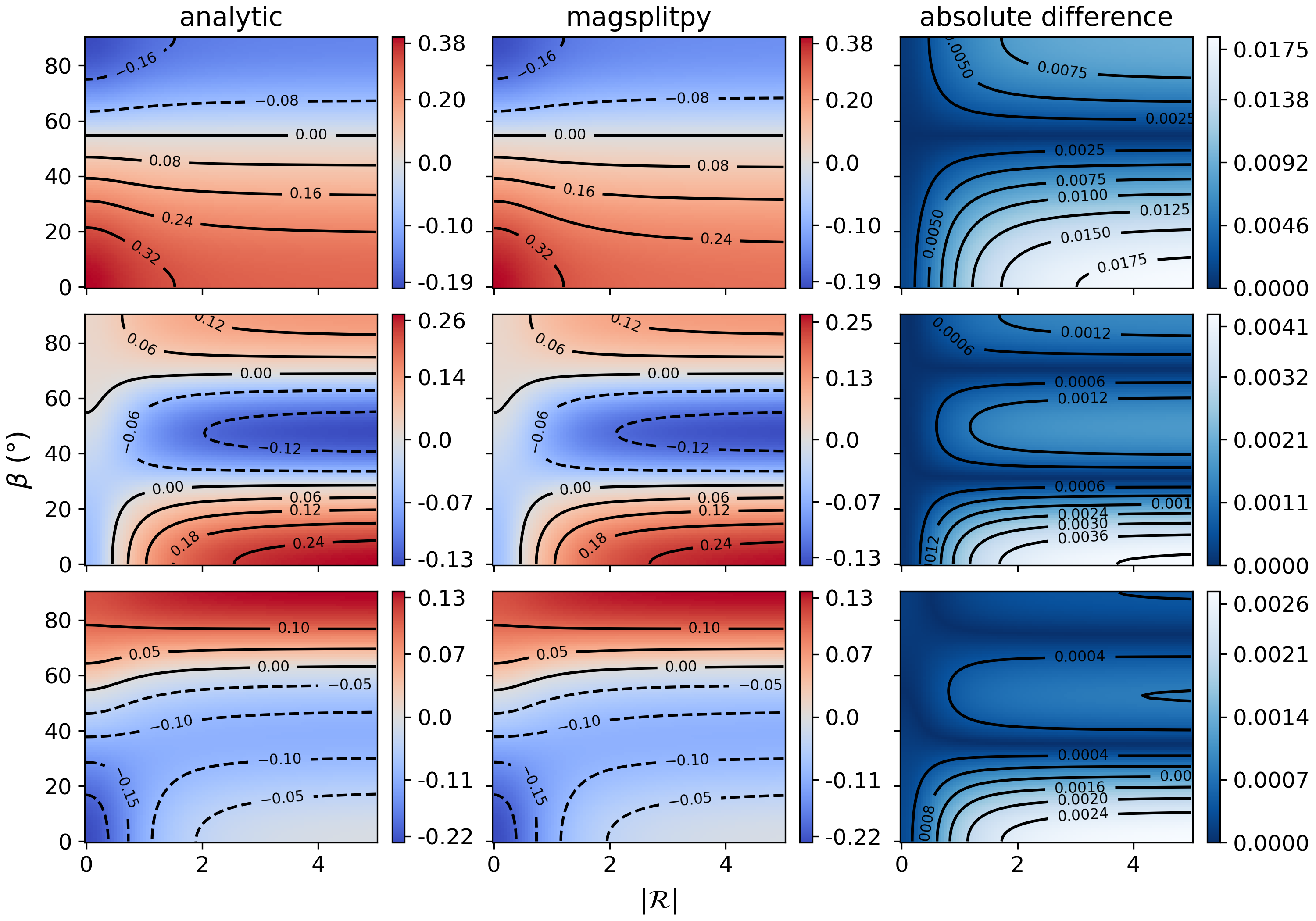}
    \caption{Colormaps for each asymmetry parameter (top: $a_{11}$, middle: $a_{21}$, bottom: $a_{22}$) for a co-aligned quadrudipole $(\beta_d = \beta_q = \beta)$. The solid (dashed) overplotted contours emphasize the values of the asymmetry parameters. Only $|\cR|$ are shown since $a_{\ell|m|}$ are functions of $\cR^2$. There is a symmetry relation of $a_{\ell|m|}(90\degree - \beta) = a_{\ell|m|}(90\degree + \beta)$ intrinsic to Eqs.~\ref{eqn: a11}-\ref{eqn: a22}, therefore we limit $\beta$ to the interval $[0\degree, 90\degree]$. Left panels: $\ell=1,2$ theoretical degeneracy for a magnetic field of the case (a) in Fig. \ref{fig:field_config_panel}. Middle panels: same degeneracy computed with \texttt{magsplitpy}. Right panels: Absolute difference between the numerical calculations and the analytical expressions of the asymmetry parameters.}

    \label{fig:magsplitpy_params_aligned}
\end{figure*}
% \begin{figure}[ht]
%     \centering
%     \includegraphics[width=0.44\textwidth]{Figures/aellm_colaigned.pdf}
%     \caption{$\ell=1,2$ degeneracy colormaps for each asymmetry parameter or co-aligned quadrudipole $(\beta_d = \beta_q = \beta)$. The solid (dashed) overplotted contours refer to positive (negative) values of asymmetry parameters. Only the positive values of $\cR$ shown since $a_{\ell,|m|}$ are functions of $\cR^2$.}
%     \label{fig:a_contours}
% \end{figure}

We first explore the simpler Case (a) of an aligned dipole and quadrupole defined in Eq.~\ref{eq:aligned}, i.e., $\beta_d = \beta_q = \beta$, as previously done for $\ell=1$ oscillation modes in \cite{Mathis2023}. This reduces the parameters describing the topologies from the previous section to $(\cR, \, \beta)$. 

\subsubsection{Analytical results from the asymmetry parameters associated with the radial component of the field} \label{sec:analytical_asym}

Plugging in the analytical definition of $B_r$ from Eq.~\ref{eqn: quadrudipole_Br_expanded} into the simplified analytic expression in Eq.~\ref{eqn: M_analytical} and the definition of asymmetry parameter in Eq.~\ref{eqn: asymmetry_parameter}, we obtain
\begin{equation} \label{eqn: a11}
    a_{11} = \frac{(7 + 5\cR^2)(1 + 3\cos{2\beta})}{70(1+\cR^2)} \, ,
\end{equation}
\noindent for $\ell=1$ modes as in \cite{Mathis2023}, and
\begin{align}
    a_{21} &= \frac{-2 + 5\cR^2 + 2(- 3 + 5 \, \cR^2) \cos{2\beta} + 25\cR^2 \cos{4\beta}}{140 \left(1+\cR^2\right)} \, ,\label{eqn: a21}\\
    a_{22} &= - \frac{16 + 5\cR^2 + 4(12 + 5 \, \cR^2) \cos{2\beta} - 25\cR^2 \cos{4\beta}}{280 \left(1+\cR^2\right)} \, \label{eqn: a22}
\end{align}
\noindent for $\ell=2$ modes. Our study only considers $\beta\in[0\degree:90\degree]$ for all three $a_{\ell|m|}$ because of the symmetry relation
\begin{equation}\label{eq:sym_relation}
    a_{\ell|m|}(\cR, 90\degree-\beta) = a_{\ell|m|}(\cR, 90\degree+\beta).
\end{equation}

The first column of Fig.~\ref{fig:magsplitpy_params_aligned} represents the value of the $a_{\ell |m|}$ parameters as function of $\beta$ and $\cR$, from Eqs.~\ref{eqn: a11}, \ref{eqn: a21} and \ref{eqn: a22}, where only the radial component of the magnetic field is used. %From it, we can explore if we should expect theoretical degeneracies when using $a_{11}, a_{21}$ and $a_{22}$ to infer $(\cR, \, \beta)$.
The three rows in Fig.~\ref{fig:magsplitpy_params_aligned} show $a_{11}$, $a_{21}$ and $a_{22}$ (from top to bottom), respectively. 
Contour maps in these left panels of Fig.~\ref{fig:magsplitpy_params_aligned} can be used to show that there are no theoretical degeneracies between the quadrudipole and a pure dipole when using all three asymmetry parameters simultaneously. We see that if, from observations, we get $a_{11} > 0$, $a_{21} > 0$ and $a_{22} < 0$, our possibility of configurations is limited to a strong quadrupole with a low inclination with respect to the rotation axis. A similar visual analysis of Fig.~\ref{fig:magsplitpy_params_aligned} shows how the availability of these three asymmetry parameters drastically reduces the possibilities in magnetic configurations. We conclude that the degeneracy of the quadrudipole with a pure dipolar field observed in \cite{Mathis2023} is lifted when accessing $\ell=1$ and $\ell=2$ oscillation frequencies simultaneously. We discuss this in detail in Section~\ref{sec:lift}.

\begin{figure*}[ht]
    \centering
    \includegraphics[width=0.94\textwidth]{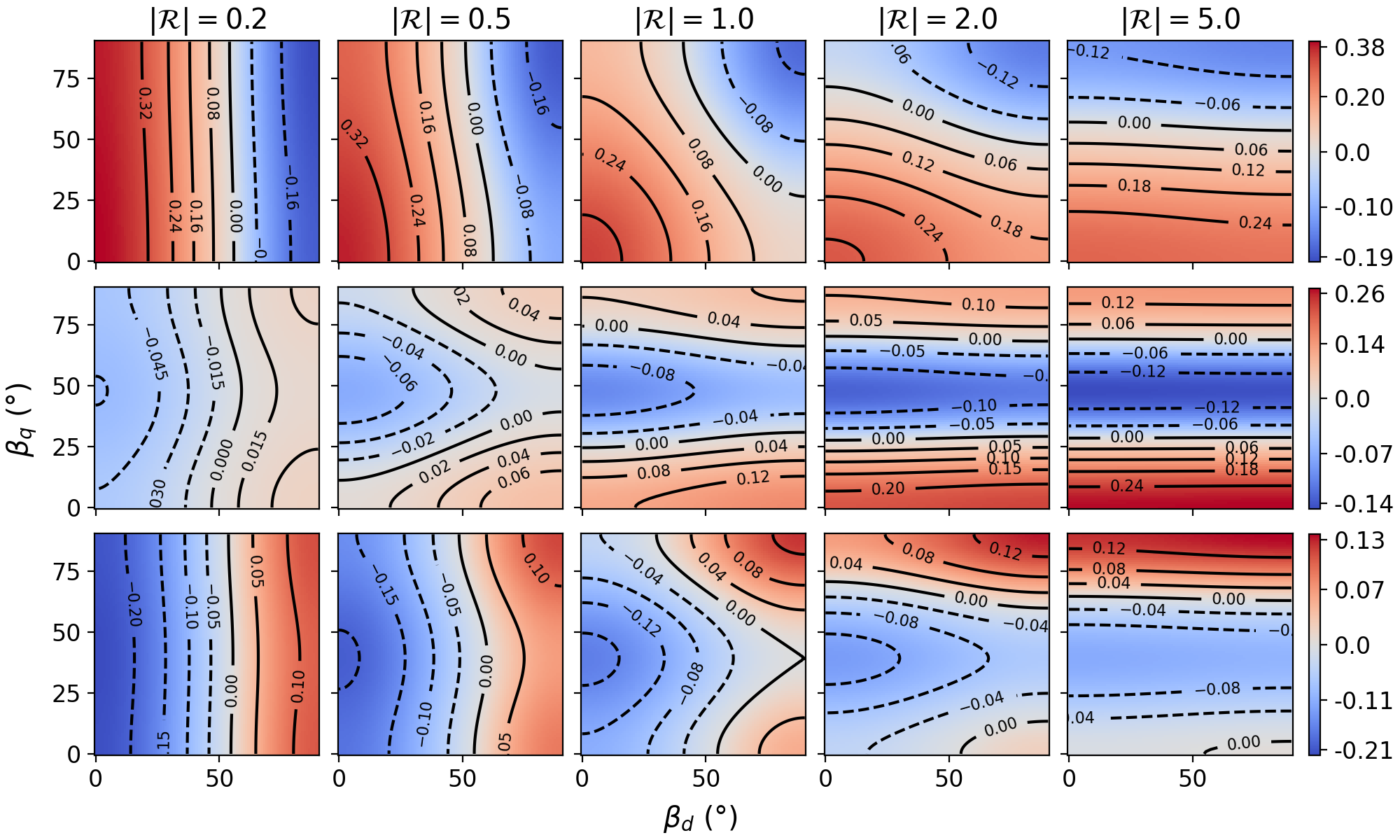}
    \caption{Colormaps for each theoretical asymmetry parameter (top: $a_{11}$, middle: $a_{21},$ bottom: $a_{22}$) for different values of $\cR$. The solid (dashed) overplotted contours emphasize the values of the asymmetry parameters. Only the positive values of $\cR$ are shown since $a_{\ell|m|}$ are functions of $\cR^2$. There are symmetry relations of $a_{\ell|m|}(\cR, 90\degree-\beta_d, \beta_q) = a_{\ell|m|}(\cR, 90\degree+\beta_d, \beta_q), a_{\ell|m|}(\cR, \beta_d, 90\degree-\beta_q) = a_{\ell|m|}(\cR, \beta_d, 90\degree+\beta_q)$ intrinsic to Eqs.~\ref{eqn: a11_full}-\ref{eqn: a22_full}, therefore we limit $\beta_d$ and $\beta_q$ to the interval $[0\degree, 90\degree]$. Shows from left to right how the theoretical asymmetries vary with increasing $|\cR|$.}
    \label{fig:analytic_params_misaligned}
\end{figure*}

We would also like to point out some salient features of the aligned quadrudipole: (i) For $a_{11}$, we recover the null line in asymmetry at $\sim 54.7^{\circ}$ consistent with the findings in \cite{Mathis2023}. For all magnetic obliquity below this angle, $a_{11}$ is positive and vice-versa. For a given $\beta$, the strength of the asymmetry increases for a stronger dipolar component (smaller $|\cR|$), (ii) For a pure dipole (or a small quadrupolar component) $a_{21}$ goes from being positive for high magnetic inclination to negative for intermediate and low inclinations. This changes to a double positive lobe at low and high $\beta$ and a negative dip in intermediate $\beta$ for stronger quadrupolar contribution. (iii) For $a_{22}$ we once, again have a more two-sided polarity where for low $\beta$ the asymmetry is negative and vice-versa. The null line has a qualitatively different trend than $a_{11}$. Further, for very low inclinations and very strong quadrupolar contribution, we also have a near zero $a_{22}$.

% Continuing along the lines of the discussion on degeneracy breaking after Eq.~\ref{eqn: a22}, we see that if, from observations, we get $a_{11} > 0$, $a_{21} > 0$ and $a_{22} < 0$, our possibility of configurations is limited to a strong quadrupole with a low inclination with respect to the rotation axis. A similar visual analysis of Fig.~\ref{fig:magsplitpy_params_aligned} shows how the availability of these three asymmetry parameters drastically reduces the possibilities in magnetic configurations. We conclude that the degeneracy of the quadrudipole with a pure dipolar field observed in \cite{Mathis2023} is lifted when accessing $\ell=1$ and $\ell=2$ oscillation frequencies simultaneously.

\subsubsection{Numerical calculation for the 3D magnetic field} \label{sec:numerical_params_aligned}

To be able to recover the signature of the full 3D magnetic field, we use \texttt{magsplitpy}. We use the magnetic field configuration of aligned quadrudipole as shown in Case (a) of Fig. \ref{fig:field_config_panel}. We calculate the magnetic splitting of these fields for the modes $(n=-52, \, \ell=1)$ and $(n=-95, \, \ell=2)$. These are obtained from the diagonal elements of the matrix ${\bf M}_{\ell}$ constructed according to Eq.~\ref{eqn:lamda_decomp}. Even though we do not explicitly compute the rotational elements in the matrix, the underlying assumption here is that rotational effects dominate magnetic effects for the class of red giants we are interested in. This renders the matrix diagonally dominant \citep[see supplementary section in][]{Li2022}. The computation of asymmetry parameters (which involves computing the asymmetric splitting and the trace of the matrix) is independent of rotational effects for a slow rotator. The above considerations allow us to simplify our analysis in this study by requiring the computation of only the magnetic part of the coupling matrix $\textbf{M}_{\ell}$. Finally, for the aligned quadrudipole, we compute the asymmetry parameters following Eq.~\ref{eqn: asymmetry_parameter} for a range of values of magnetic inclination $\beta$ and quadrupolar contribution $|\cR|$. 

 % \subsubsection{Result comparison}
Fig.~\ref{fig:magsplitpy_params_aligned} also compares the numerical results obtained using $\texttt{magsplitpy}$ with the analytical relations found above. The middle panel shows the numerical results implementing the full 3D vector field. For the same asymmetry parameter, we have used the same color bar for both the analytical and numerical results for ease of comparison. The rightmost panel shows the absolute difference between the analytical and numerical results. The difference between the analytical and numerical results is very promising: for $a_{11}$ the maximum error is around 5\% while for $a_{21}$ and $a_{22}$ its around 1\%. {This comparison demonstrates that the approximation leading to the theoretical expressions for asymmetry parameters in our study and in \cite{Mathis2023} are valid. For simple magnetic field geometry such as Case (a), theoretical expressions can be used with confidence to relate $(a_{11}, a_{21}, a_{22})$ to $(|\cR|, \beta)$.} 

% \sout{Both of these are much smaller than the error in asymmetry parameters (estimated to be around $\pm 0.08$ in Appendix.~\ref{sec: a_uncertainty}.}\textcolor{red}{We don't know that yet, we can explain it is the uncertainty section instead? See new paragraph in section 4.2} \textbf{Okay, I agree.} 

%\sout{However, we prove here that, even if magnetic field topologies are too complex for analytic developments, \texttt{magsplitpy} is always capable of producing the link between $(a_{11}, a_{21}, a_{22})$ and $(\cR, \beta)$.}}
%\textbf{I think I wouldn't link it to  $(\cR, \beta)$ because for complex fields, we might not be able to use these as parameters. Instead, I would frame the above sentence as 
{This theoretical benchmarking also proves that \texttt{magsplitpy} offers a general numerical framework to reliably compute $(a_{11}, a_{21}, a_{22})$ even if magnetic field topologies are too complex for analytic developments. This has important implications in terms of setting up an inverse problem, for instance when using Bayesian inference schemes. To further demonstrate the potential of \texttt{magsplitpy}, we also present the results benchmarking against the analytical results for a misaligned quadrudipole (see Section~\ref{sec:misaligned} and Appendix~\ref{sec:numerical_misaligned}).}

\subsection{{Non-Aligned rotation, dipolar field, and quadrupolar field axes}}
\label{sec:misaligned}
{Section~\ref{sec:aligned_quadrudipole} investigated the simplest case of a multi-moment $\Bv$ field for an aligned quadrudipole. In this section, we explore the degeneracies of Case (b), a misaligned quadrudipole where the dipolar and quadrupolar components have different inclination angles with respect to the rotation axis. In the case of a misaligned quadrudipole, 
\begin{equation} \label{eqn: a11_full}
    a_{11} = \frac{7 + 5\cR^2 + 21 \cos{2\beta_d} + 15 \, \cR^2 \cos{2\beta_q}}{70(1+\cR^2)} \, .
\end{equation}
Again, we note the following specific cases (i) for $\cR = 0$ and $\beta_d = \beta_q = \beta$, this matches with Eq.~26 of \cite{Mathis2023}, and (ii) for $\beta_d = \beta_q = 0$, this matches with Eq.~29 of \cite{Mathis2023}. Following the same method, we find that for $\ell=2$, the two asymmetry parameters take the following expressions:
\begin{align}
    a_{21} &= \frac{-2 + 5\cR^2 - 6\cos{2\beta_d} + 10 \, \cR^2 \cos{2\beta_q} + 25\cR^2 \cos{4\beta_q}}{140 \left(1+\cR^2\right)} \label{eqn: a21_full}\, ,\\
    a_{22} &= - \frac{16 + 5\cR^2 + 48\cos{2\beta_d} + 20 \, \cR^2 \cos{2\beta_q} - 25\cR^2 \cos{4\beta_q}}{280 \left(1+\cR^2\right)} \label{eqn: a22_full}\, .
\end{align}
Similar to Eqs.~\ref{eqn: a11}-\ref{eqn: a22} there is a symmetry relation present on both angles in Eqs.~\ref{eqn: a11_full}-\ref{eqn: a22_full}. Namely,
\begin{align}\label{eq:sym_relation_full}
    a_{\ell|m|}(\cR, 90\degree-\beta_d, \beta_q) &= a_{\ell|m|}(\cR, 90\degree+\beta_d, \beta_q),\\
    a_{\ell|m|}(\cR, \beta_d, 90\degree-\beta_q) &= a_{\ell|m|}(\cR, \beta_d, 90\degree+\beta_q).
\end{align}

In Appendix~\ref{sec:numerical_misaligned} we study the particular case of $\beta_q - \beta_d = 90\degree$. We represent on Fig.~\ref{fig:magsplitpy_params_misaligned} the values of the asymmetry parameters, calculated from the analytical formula and with \texttt{magsplitpy}, depending on the ratio of the field amplitudes at the H-shell and of the inclination $\beta_d$. As in Case (a), the analytical expressions provide robust results with similar precision compared to the \texttt{magsplitpy} results. % on the left column of Fig.~\ref{fig:magsplitpy_misaligned}. 
We observe that, as the quadrupolar field strength increases with respect to the dipolar strength ($|\cR|$ increases), the variation of the asymmetry parameter values compared to Case (a) increases. As this effect must also depend on the angle $\beta_q$, we variate the three parameters ($|\cR|, \beta_d, \beta_q)$ and represent the results in Fig.~\ref{fig:3d_projection_2} (see discussion in Appendix~\ref{sec:numerical_misaligned}).

% \textcolor{purple}{Lukas: this is your paragraph, rewrite things the way it makes sense to you to describe Fig 4!} \sout{\textcolor{blue}{Needs to be discussed}
% When changing the $\beta_q-\beta_d$ angle, the asymmetry colormap gets bent. At a fixed |R| ratio, increasing $\beta_q-\beta_d$ results in a shift of the colormap to higher/lower \textcolor{blue}{@Lukas which one?} $\beta$ angles. The higher the |R|, the larger the absolute shift. As a result, there is a complete degeneracy between topology cases (a) and (b), and one cannot say whether the quadrupolar component of the field is aligned or not with the dipolar component from asymmetry measurement. This degeneracy is represented on Fig.~\ref{fig:??} \textcolor{blue}{New plot from Lukas}}

{To summarize, Fig.~\ref{fig:analytic_params_misaligned} shows how the value of $|\cR|$ affects the asymmetry parameters. For $|\cR| = 0.2$ the contour lines are almost vertical, indicating that $a_{\ell|m|}$ are independent of the quadrupole angle $\beta_q$ (reasonable since the dipolar field dominates). For $|\cR| = 5.0$ the result is the opposite, but there the asymmetry parameters are independent of the dipole angle. Both $a_{11}$ and $a_{22}$ slowly interpolate between those two extrema as the field goes from dipole-dominated to quadrupole-dominated ($|\cR|$ increases) while the magnitude of the asymmetry parameters stays roughly constant. The parameter $a_{21}$ is more sensitive to the quadrupolar component of the field than $a_{11}$ and $a_{22}$. For small $|\cR|$ we have $a_{21} \approx 10^{-2}$ which grows by an order of magnitude as $|\cR| \to 1$. Thus, the magnitude of $a_{21}$ is %\sout{\textcolor{red}{the most reliable way to confine the degeneracy in}} 
a good proxy for $|\cR|$, and constraints the quadrupolar angle.
%\sout{The parameter $a_{21}$ is more sensitive to the quadrupolar component of the field than $a_{11}$ and $a_{22}$ and is, already at $\cR=1.0$, almost independent of the dipolar inclination.}\textcolor{red}{It is not really true, because the variation of the values of a21 are much smaller than of a11 and a22 at small R} }

\subsection{{Offset magnetic field}} 
\label{sec:offset}

Lastly, we show the signature of an offset dipolar field (Case (c) in Fig. \ref{fig:field_config_panel}) on the splittings of the $\ell=1$ and $\ell=2$ oscillation modes, as observed at the surface of stars with radiative envelopes \citep{Donati2009, Vennes2017, Hardy2023, Hardy2023a, Hollands2023}. For this scenario, we directly use the numerical method, as the theoretical development of the asymmetry parameters becomes too convoluted for a proper derivation. The \texttt{magsplitpy} results are presented in Fig.~\ref{fig:magsplitpy_asyms_offset} as a function of the offset $z_o$. {We observe that, starting from a centered dipole, as we gradually offset the center of the field (increase $z_o$)} up to the radius of the H-shell, all $|a_{\ell|m|}|$ converge to zero. This is because the magnetic field topology probed at the H-shell varies with $z_o$, going from purely dipolar to higher-order components when $z_o$ increases. As a result, the magnetic field averages out along the H-shell, resulting in null asymmetries. Once $z_o$ is greater than $R_h$, there are no longer {radial magnetic field lines of opposite sign} canceling out at the H-shell. 
%\textbf{[NOT SURE IF I AGREE WITH THIS SINCE WE ARE MOSTLY PROBING $B_r^2$. Indeed, should we phrase it as: there are no longer radial magnetic field components of opposite sign canceling out at the H-shell]?}
As a result, $|a_{\ell|m|}|$ values increase, even though the field probed is of low amplitude {(this increase depends on the geometry of the field, and might vary with the choice of the radial profile)}.
%\textcolor{orange}{(This is not necessarily the case/ depends on the radial dependence. My guess of why this happens would be, that the asymmetry of an "$\ell=0$"-field is smaller than one of a dipole.)} 
Eventually, as $z_o$ increases, the magnetic field amplitude at the H-shell decreases, and all $|a_{\ell|m|}|$ converge to zero again. We demonstrate through the model Case (c) that a small offset of a large-scale magnetic field along the rotation axis of about $3\%$ of the extent of the radiative cavity leads to a disappearance of magnetic field effects on the symmetry of all the modes, due to the geometry of the H-shell. As a result, large-scale magnetic fields can have the same effect as small-scale magnetic fields \cite[such as the one resulting from dynamo action, see e.g.][]{Fuller2019, Petitdemange2023}}.
\begin{figure}[t]
    \centering
    \includegraphics[width=0.47\textwidth]{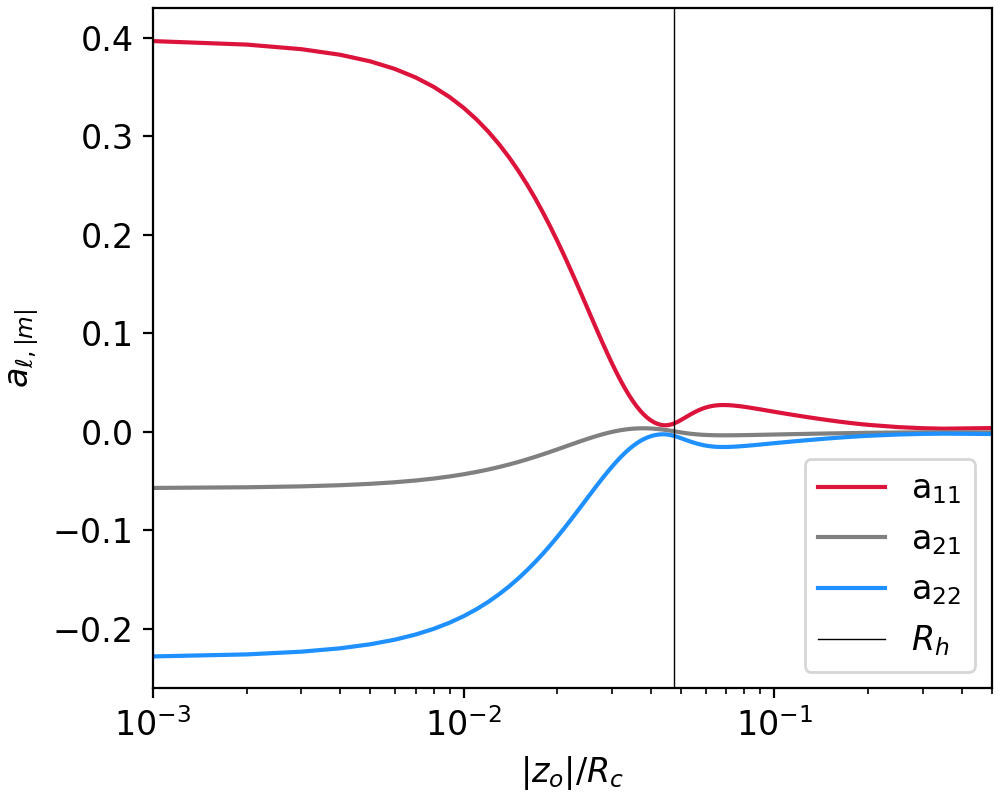}
    \caption{Asymmetry parameters $a_{\ell,|m|}$ depending on the offset of a magnetic field of type (c) in Fig. \ref{fig:field_config_panel}. The offset $z_o$ of the field is given relative to the radiative zone radius. The black dashed line indicates the location of the H-shell radius $R_h$.}
    \label{fig:magsplitpy_asyms_offset}
\end{figure}

\begin{figure*}[t]
    \centering
    \includegraphics[width=0.7\linewidth]{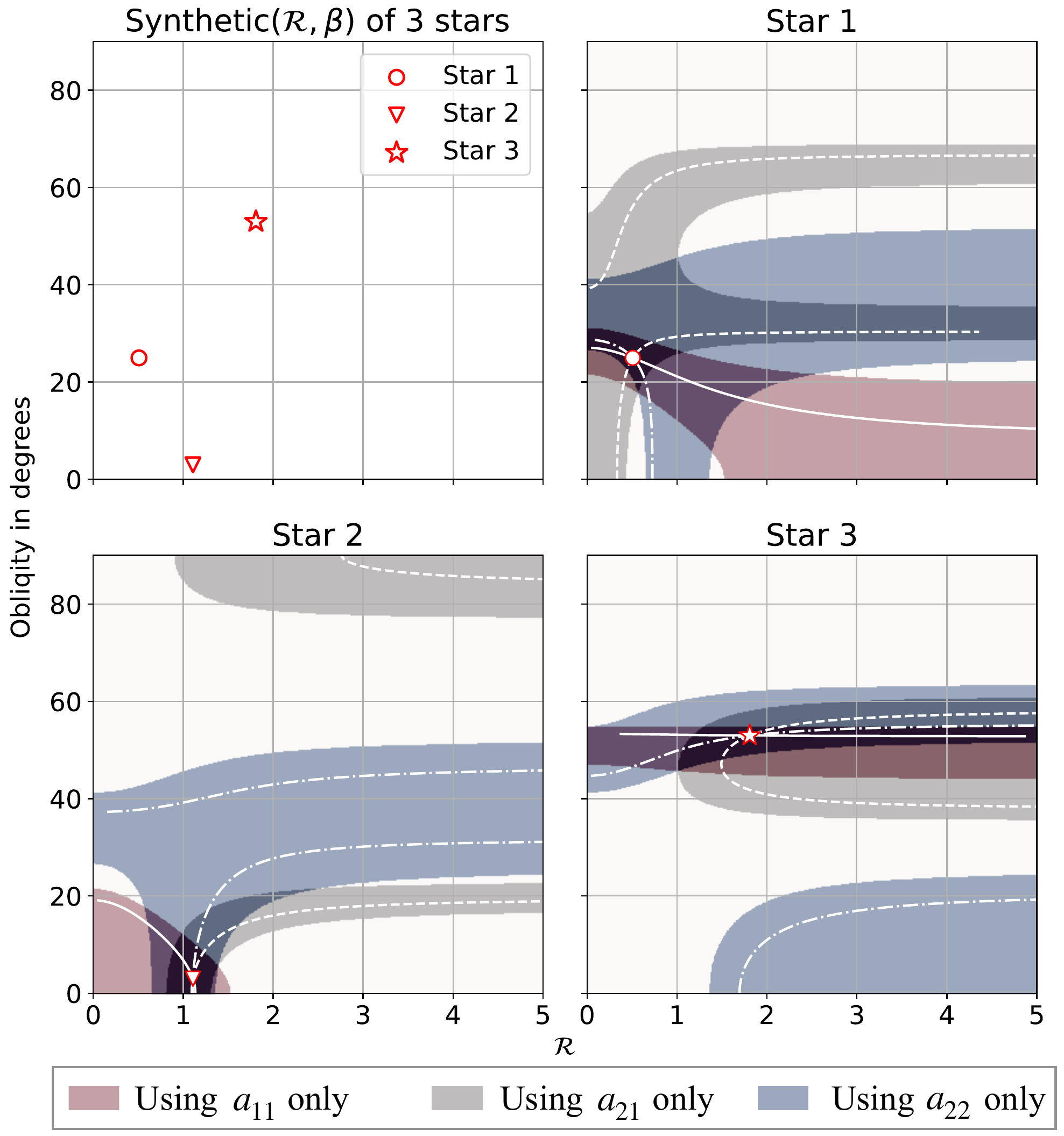}
    \caption{Overlapping bands of iso-asymmetry values across $a_{11}, a_{21}$ and $a_{22}$ demonstrating the drastic reduction of degeneracy in $(|\cR|, \beta)$ space for a co-aligned quadrudipole when using $\ell=1$ and $\ell=2$ oscillation modes simultaneously. The resp. white line, dashed line, and dashed-dotted line indicate the analytical solution corresponding to the measurement of resp $a_{11}, a_{21}$ and $a_{22}$ from Section~\ref{sec:aligned_quadrudipole}. The width of colored bands indicating typical uncertainty on the measurement of the asymmetry parameters is chosen to account for the finite data frequency resolution of Kepler 4 years following Section~\ref{sec:resolution}. }
    %\textcolor{red}{Could you put the lines in white instead of black so that we can see them even in the black region? And then the symbols would be black+white outline instead of white. + change associated text and legend} \textbf{Done. Black outline was not visible in the dark regions. So I made it red after playing around a bit.}}
    \label{fig:3star_degeneracy}
\end{figure*}

\section{Results \& Discussion}
\label{sec:discussion}

 \subsection{Detectability of quadrudipole magnetic fields from $\ell=1,2$ modes}
\label{sec:lift}

% \subsection{Lifting topology degeneracies from $\ell=2$ modes}
From the results in Section~\ref{sec:probe}, we demonstrate the detectability of quadrupolar magnetic field components from combined measurement on the $\ell=1$ and $\ell=2$ mixed mode frequencies. The top left panel of Fig.~\ref{fig:3star_degeneracy} presents three random examples of $(\cR, \beta)$ configuration in the case (a) of an aligned quadrudipole. The corresponding three asymmetry parameters are calculated, and their possible values to properly recover $(\cR, \beta)$ are represented in the $\beta$ versus $|\mathcal{R}$| diagram in the other three panels. In the first step, we demarcate (with white lines) the exact degeneracies in $(\cR, \beta)$ when using measurement only one kind of asymmetry parameter (either $a_{11}$ or $a_{21}$ or $a_{22}$). The shaded areas around the lines of single $a_{\ell |m|}$ degeneracy represent typical uncertainties which we will analyze in Section~\ref{sec:resolution}.

For Star 1, taking one asymmetry parameter only (such as using only $a_{11}$ from measuring the frequency splittings in $\ell=1$ modes) leads to a total degeneracy of the quadrudipole with a purely dipolar field \citep[as demonstrated by][]{Mathis2023}. Adding a second asymmetry parameter (i.e. using $\ell=2$ modes) completely lifts this degeneracy, and both $\beta$ and $|\cR|$ can be measured. Using the three asymmetry parameters confirms the measurement and allows us to extract simultaneously exact measurements of $\beta$ and $|\mathcal{R}$|. The same conclusion can be drawn for Star 2 if asymmetry parameters are known with high precision (see Section~\ref{sec:resolution} for more details about uncertainties on asymmetry parameters). However, a null asymmetry parameter, as is the case for $a_{11}$ in the case of Star 3 where $\beta\approx 55\degree$, is the worst-case scenario for a characterization of the magnetic field topology. Indeed, any small-scale magnetic field that averages out in the H-shell, or a non-magnetic radiative zone, results in a null asymmetry parameter. It is therefore much harder to constrain the magnetic field topology for field inclination angle nearing $55\degree$, as illustrated on the bottom right panel of Fig.~\ref{fig:3star_degeneracy}.

\begin{figure}[ht]
    \centering
    \includegraphics[width=0.47\textwidth]{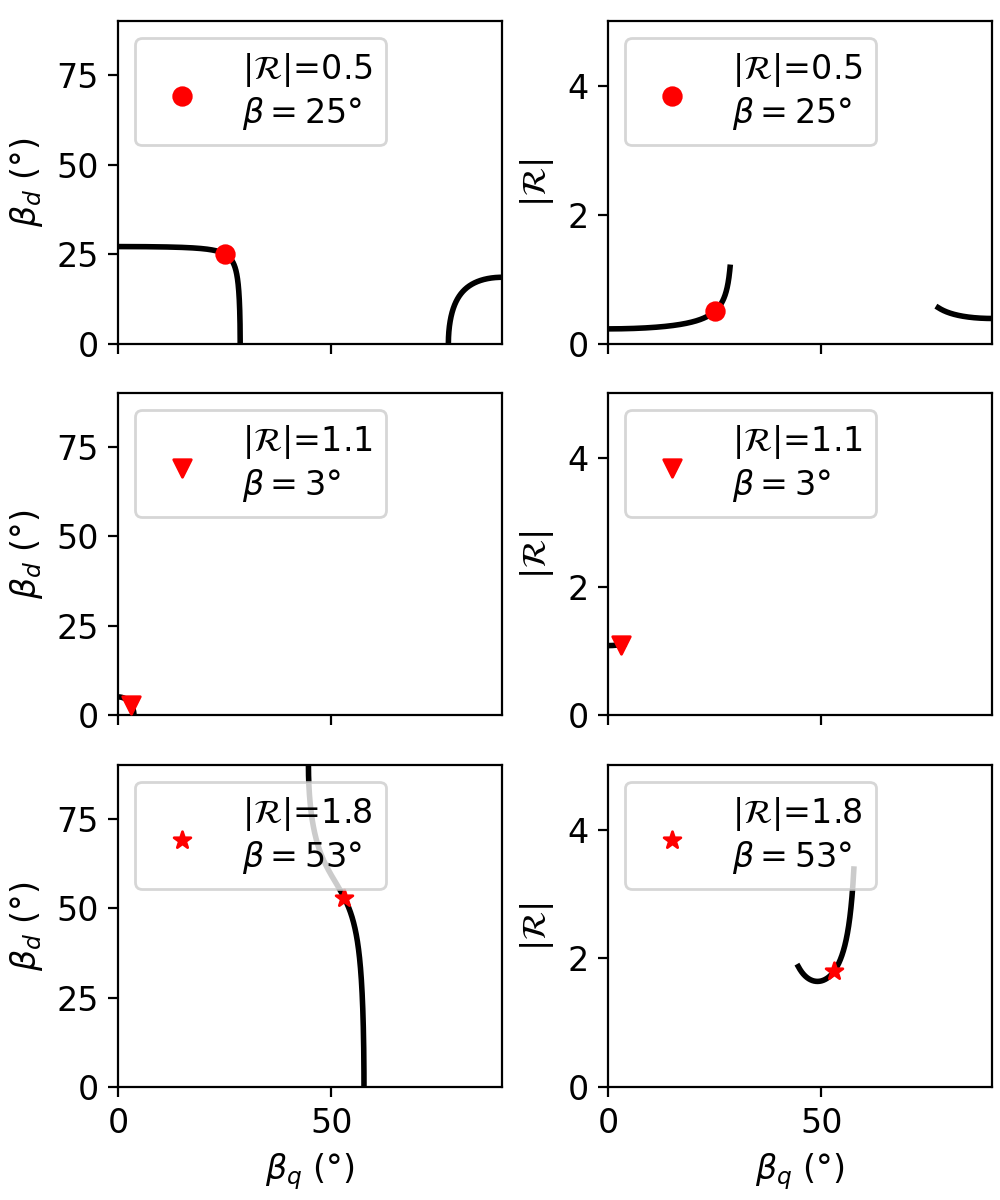}
    \caption{Degeneracies in the asymmetry parameters for the three stars on Fig.~\ref{fig:3star_degeneracy}. Each column is a projection on the plane of two of the three field topology parameters $(\beta_q, \beta_d)$, $(\beta_q, \cR)$. 
    %\textcolor{red}{matcal missing on the y-axis}
    }
    \label{fig:3d_projection_degeneracy_1}
\end{figure}

\subsection{Impact of the frequency resolution in the data on the degeneracy}
\label{sec:resolution}

As every asteroseismic measurement comes with its own uncertainty, we add on Fig.~\ref{fig:3star_degeneracy} the effect of the uncertainty on the measure of the asymmetry parameters on the inversion of $\beta$ and $|\mathcal{R}$| through the shaded regions. The uncertainty on asymmetry parameters is calculated as (see Appendix~\ref{sec: a_uncertainty}):
\begin{equation}
\label{eq:uncertainty}
    \delta a_{\ell|m|} \approx \frac{\sqrt{6} \, \delta f}{\delta \omega_{B}^{\ell}}
\end{equation}
\noindent with $\delta f$ the frequency resolution in the data and $\delta \omega_{B}^{\ell}$ the averaged magnetic shift modulated by the $\zeta$ function as defined in Appendix~\ref{sec: a_uncertainty}. 
%\textcolor{blue}{@Srijan check and modify if necessary to match with the appendix} %It %demonstrates how, using the three measures of asymmetry with associated uncertainty, we can severely narrow down the range of $\beta$ and $\cR$ of the RGB fossil field. 
%In this synthetic exercise, we randomly choose three \textit{true} values of $(\cR, \beta)$ indicating three stars . 
For each of the chosen model stars in section~\ref{sec:lift}, we show patches of light, intermediate, and dark shades. The light-shaded area is indicative of the abundance of degenerate $(|\cR|, \beta)$ configurations on the availability of only one asymmetry parameter. The intermediate shaded region shows the reduced degeneracy constrained by using two asymmetry parameters. The darkest shade, which is the smallest patch, shows the restricted area in $(|\cR|, \beta)$ space when all three asymmetry parameters are available. Note that the shaded area increases with increasing data frequency resolution, here taken as $8$nHz to mimic \textsl{Kepler} 4-year data resolution. As a result, the quality of data of course plays a major role in the detectability of complex magnetic field topology. Typical \textsl{Kepler} data uncertainty leads to a small degeneracy for Star 1, where the presence of a small quadrupolar component becomes debatable. The inclination angle $\beta$ remains well-constrained. For Star 2, both the presence of the quadrupolar component and the inclination angle of the field remain well-constrained even with uncertainties on the asymmetry parameters. If Star 3 hosts a quadrudipole, it surely has a strong quadrupolar component and a well-constrained inclination angle near $55\degree$, but the quadrupole to dipole strength ratio cannot be well constrained. This goes back to the discussion in Section~\ref{sec:lift} about near-zero asymmetry parameter values. In short, depending on the true $(|\cR|, \beta)$ parameters, the degeneracy regions can vary, and some stars might be more easily characterized than others.

{The uncertainty on the $a_{\ell |m|}$ parameters from Eq.~\ref{eq:uncertainty} is about 0.08 when taking $\omega_B^\ell=0.2\mu$Hz.
%\textcolor{red}{we say 0.2 in the appendix, which one is it?} \textbf{Not sure why it was 0.02. Should have been 0.08 as per Appendix}. 
For $|a_{\ell|m|}|$ ranging from 0 up to 0.4, the minimum uncertainty associated with observational constraints with the \textsl{Kepler} mission is about $5\%$. As we demonstrated in Section~\ref{sec:numerical_params_aligned}, the error induced by the chosen methodology to compute asymmetry parameters is lower than $5\%$ {for $a_{11}$ and lower than $1\%$ for $a_{21}$ and $a_{22}$}, the dominant source of uncertainty on $(|\cR|, \beta)$ results indeed from observations and not from the chosen methodology.} 
%{\color{blue}Re-make the plot for an appropriate frequency resolution, add the exact solution (see Lukas plot), and connect it with Kepler and PLATO data.}

% \begin{table*}[ht]
% \centering
% \caption{Approximated access to asymmetry parameters depending on the inclination of the star with the line of sight, following mode visibilities as in \cite{Gizon2003}.}
% \begin{tabular}{ c || c| c| c| c| c |}
% % \hline\hline
%  & $0\lessapprox i\lessapprox 10 \degree$ & $10 \lessapprox i\lessapprox 40 \degree$ & $40\lessapprox i\lessapprox 70\degree$ & $70 \lessapprox  i\lessapprox 80 \degree$ & $80 \lessapprox i\lessapprox 90 \degree$\\ [0.5ex] % inserts table %heading
% \hline 
% $a_{11}$& \textcolor{gray}{\xmark} & \cmark & \cmark & \cmark & \textcolor{gray}{\xmark}\\
% $a_{21}$& \textcolor{gray}{\xmark} & \cmark & \textcolor{gray}{\xmark} & \cmark & \textcolor{gray}{\xmark}\\
% $a_{22}$& \textcolor{gray}{\xmark} & \textcolor{gray}{\xmark} & \textcolor{gray}{\xmark} & \cmark& \cmark\\
% $a_{22} - a_{21}$& \textcolor{gray}{\xmark} & \textcolor{gray}{\xmark} & \cmark & (\cmark )& \textcolor{gray}{\xmark}\\ [1ex]
% % \hline
% \end{tabular}
% \label{table:asym}
% \end{table*}

\subsection{Disentangling inclined quadrupoles}
{In Case (a), two out of three $a_{\ell |m|}$ are necessary to theoretically constrain the field topology through the measure of $|\cR|$ and $\beta$. In Case (b), however, we need three independent $a_{\ell |m|}$ to fully constrain $\cR$, $\beta_d$, and $\beta_q$. As a result of Appendix~\ref{app:matrix}, we notice that $a_{\ell |m|}$ depend on each other through the relation:}
\begin{equation}
    a_{22} + a_{21} + \frac{5 \, M_{2}^{0,0}}{\mathcal{T} \, \overline{B_r^2}(a_{11})} = 1 \, .
\end{equation}
where $\overline{B_r^2}(a_{11})$ is an estimate of the horizontal average of the squared average magnetic field across the $g$-mode cavity obtained from the $a_{11}$ measurements using only $\ell=1$ modes. $\mathcal{T}$ captures the integrated effect (in radius) of the mode sensitivity across the $g$-mode cavity in the radiative interior (see Eq.~\ref{eqn: prefactor}) which is independent of field strength or topology. For a given field topology, $M_{2}^{0,0}$ takes a particular value (shift of the $m=0$ quadrupolar mode). Therefore, the above equation with three asymmetry parameters only has two degrees of freedom, i.e., the third asymmetry parameter is determined from the knowledge of the other two.

{Therefore, we can only measure two independent asymmetry parameters to constrain the field topology. The degeneracy between the quadrudipole with or without inclination of the quadrupole with respect to the dipole axis is therefore not fully lifted when using $\ell=1$ and $\ell=2$ oscillation modes. Octupolar mixed-mode frequencies would be required, which are very unlikely to be detected in current datasets. Figure~\ref{fig:3d_projection_degeneracy_1} shows the resulting degeneracies for the three stars in Fig~\ref{fig:3star_degeneracy}. For {Star 1}, assuming a quadrudipole configuration, we can deduce that {(i)} the dipole is inclined with a small angle with the rotation axis $(\beta_d \lesssim 25^{\circ})$, {(ii)} that there is at least a small quadrupolar component $\cR$, but {(iii)} the quadrupole can either be close to aligned with the dipole or close to a $90\degree$ inclination. For Star 2, the topology is much better constrained, with the dipole and quadrupole aligned with each other and {the relative strength $\cR$} close to $1$. Star 3 presents a full degeneracy in terms of the inclination angle of the dipole, while the ratio $\cR$ shows a dominant quadrupole with an axis of about $50\degree$ with respect to the dipole. Figure~\ref{fig:3d_projection_degeneracy_1} has been simplified for readability; more comprehensive 2D degeneracy maps (see Fig.~\ref{fig:3d_projection_2}) are discussed in Appendix~\ref{sec:numerical_misaligned}. Depending on the value of the asymmetry parameters, the degeneracy on the quadrudipole inclinations is therefore highly variable, and {such careful analyses} will have to be performed on a case-by-case basis.}

\begin{figure}[t]
    \centering
    \includegraphics[width=0.45\textwidth]{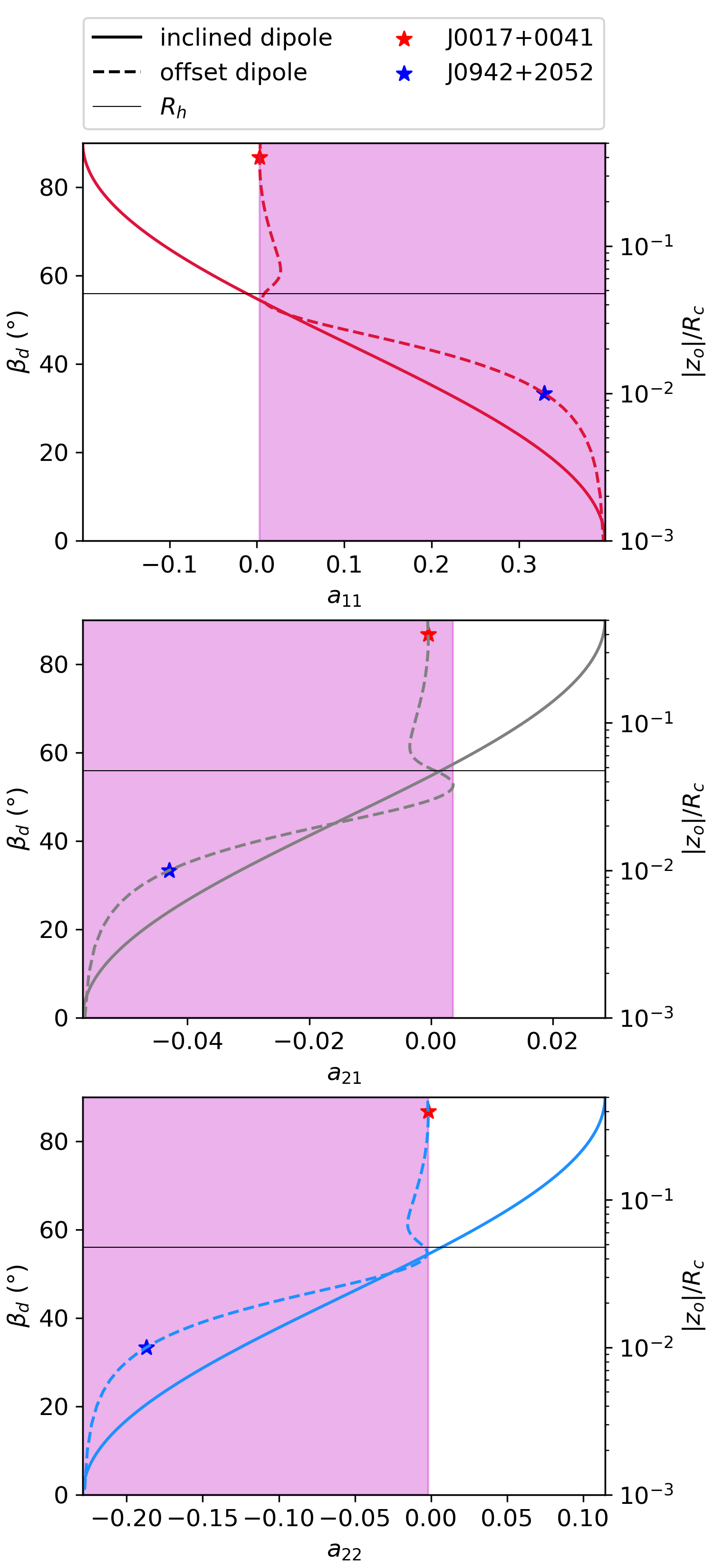}
    \caption{Asymmetry parameters $a_{\ell,|m|}$ depending on the inclination angle of the field $\beta$ in case (a) or on the offset of a magnetic field of the case (c) in Fig. \ref{fig:field_config_panel}. The offset $z_o$ of the field is given relative to the radiative zone radius. The pink region indicates, on each panel, the domain for which both an inclined dipole and an offset dipole can yield to the same asymmetry parameter. The blue and red stars show the minimum and maximum offsets measured from the combination of studies by \cite{Vennes2017, Hardy2023, Hardy2023a, Hollands2023}. J0942+2052 has an offset of -0.01 (\cite{Hardy2023a}) and J0017+004 of 0.40 (\cite{Hardy2023}).}
    \label{fig:compare_asyms_offset_inclined}
\end{figure}

\subsection{Impact of the centrality of the magnetic field on its detectability and characterization}
% \textcolor{blue}{Here discuss:
% - how off-centered field might lead to biased estimate of the magnetic field amplitude towards higher values
% - how such shifts have been observed at the surface of other stars
% - how strong off-centered (or small scale) magnetic fields can lead to a11=0 (what about the values of a21 and a22?)}

{We initiated in section~\ref{sec:offset} the discussion regarding the impact of the centrality of large-scale magnetic fields on the observed asymmetries, and therefore on the detectability of large-scale magnetic fields. In Fig.~\ref{fig:compare_asyms_offset_inclined} we present the {dependence of} asymmetry parameters {on inclination $\beta_d$ of the dipole (solid lines and left axis) and on offset $z_o$ (dashed lines and right axis)}. Even though a combination of the three asymmetry parameters is not perfectly the same in the case of inclined or the offset dipole, the solutions are very close to one another and lie within the uncertainty ranges from Appendix~\ref{sec: a_uncertainty}. {Signatures of the} inclined dipole on the symmetry of each $\ell$ multiplet can, therefore, be degenerate with those of the offset dipole for $\beta \lessapprox 55\degree$ (see the pink regions in Fig.~\ref{fig:compare_asyms_offset_inclined}).} 

%The maximum value found was $|z_o|=0.4$, and the minimum value was $|z_o|=0.01$. 
%\textcolor{purple}{There are 2 papers by Hardy in 2023, and the first one has a lot of reported az values. It doesn't change your max value, just the min value that goes to -0.01}% from a total of 10 reported white dwarfs.}
\begin{figure}[t]
    \centering
    \includegraphics[width=0.47\textwidth]{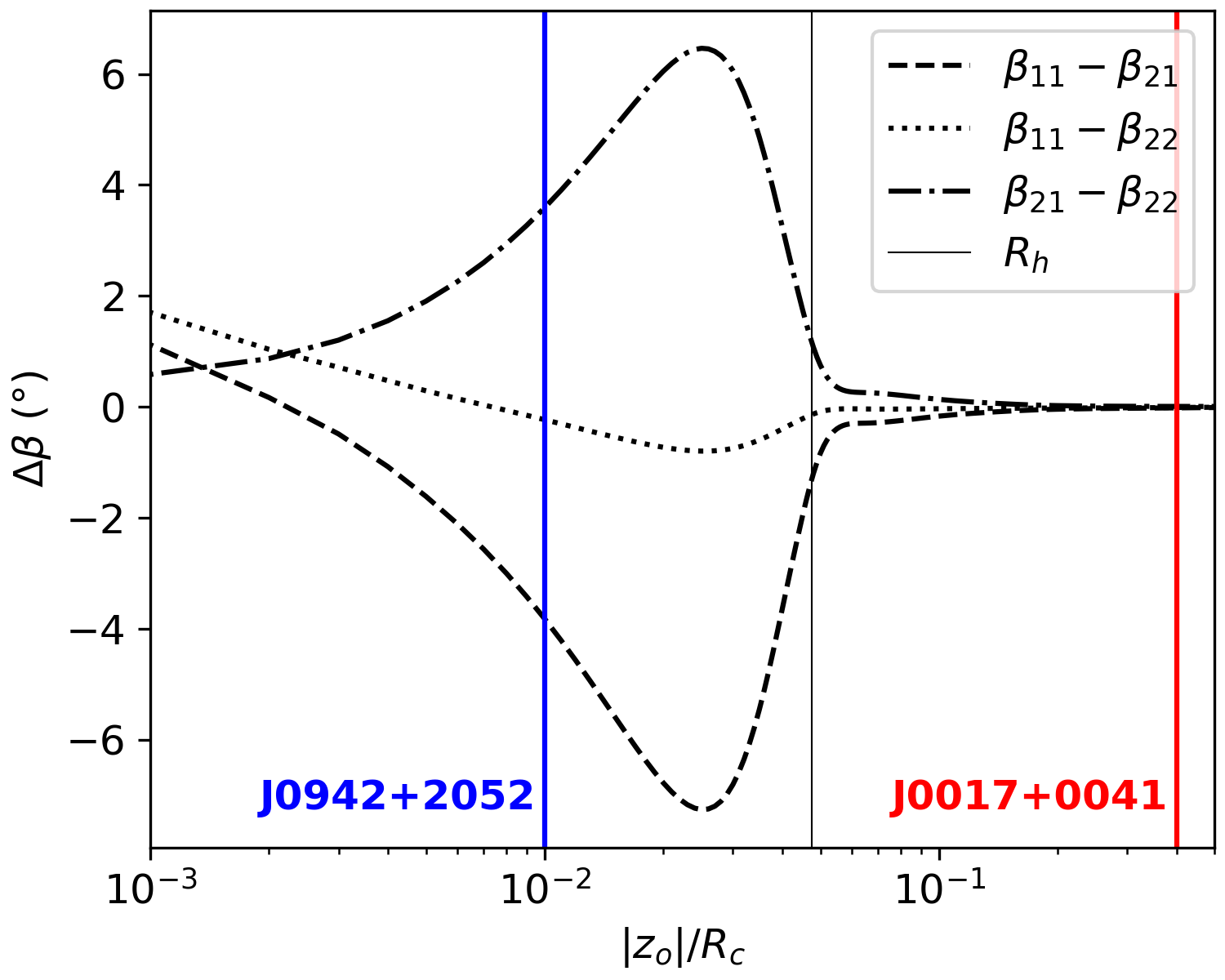}
    \caption{Differences between the inclination angles resulting from the asymmetry parameters of different offsets. }
    %For a fixed offset $z_o$ we calculate all asymmetry parameters $a_{11}(z_o), a_{21}(z_o), a_{2}(z_o)$. Then, we calculate for each asymmetry parameter the corresponding inclination angles $\beta_{\ell, |m|}$. These three inclination angles for different $a_{\ell,|m|}$ might not be the same. Thus, if the difference in angles is non zero the combination of asymmetry parameters is unique to an off-centered dipole and can be distinguished from an inclined dipole. However, the current resolution in asymmetry parameters leads to an uncertainty in the inclination angle of $\sim \pm 7°$ (see \cite{Li2022, Mathis2023} which is higher than the largest difference in angles. Thus, until the resolution in $a_{\ell, |m|}$ is increased, we cannot differentiate between off-centered and inclined dipoles. \textcolor{orange}{I put the full explanation here since I was not sure where to put it in the text.}\textcolor{purple}{Lukas the two lines do not correspond to the correct stars (it is the opposite?). Could you choose different colors than blue and red as we already have them for the l dependency? maybe orange and purple? (same on fig 8). Rh should be indicated the same way on fig 8 and 9: choose if you want to do it with continuous or dashed or dotted line.}}
    \label{fig:degeneracies_incl_offset}
\end{figure}

%\textcolor{red}{@Srijan: are the effects of inclined or off-centered fields on wB1 and wB2 the same? I think so, but just want to be sure} \textbf{I'll look into this soon.}

{Star 1 from Fig.~\ref{fig:3star_degeneracy} has asymmetry parameter values such as the field can be a quadrudipole inclined with a dipole angle of $\beta_d=25\degree$ and $|\cR|$ of about 0.5 as in Fig.~\ref{fig:3star_degeneracy}. However, a purely dipolar field with an offset along the rotation axis of about $2\%$ of the radiative zone extent leads to very similar asymmetry values (see Fig.~\ref{fig:compare_asyms_offset_inclined}). Including measurement uncertainties as described in Appendix~\ref{sec: a_uncertainty}, the offset dipole and the quadrudipole corresponding to the asymmetry parameters of Star 1 would be degenerate. For Star 2 and Star 3, quadrudipoles with the configurations discussed previously are not degenerated with an offset dipolar field, as the $a_{21}$ parameter corresponding to the quadrudipoles cannot be recovered with a dipolar field (offset or not, see Fig.~\ref{fig:magsplitpy_params_aligned}).} 
%\textbf{This paragraph is great but to be double-checked once we have the final version of Fig.~8.} \textcolor{violet}{checked and corrected}\textcolor{orange}{I don't think we can put the $\beta$ values from Fig.~\ref{fig:3star_degeneracy} into Fig. 8 since one represents the inclination of a quadrudipole and the other the inclination of a pure dipole.} \textcolor{violet}{Agreed. But we can still compare if the solutions are close, see new text. I find it interesting that the ranges for a11 and a22 for the dipole vs quadrudipole and fairly close, but the range for a21 is -0.06:0.03 for the dipole and -0.15:0.3 for the quadrudipole from fig 3.}

{Additionally, we also include upper and lower boundaries of offsets observed in white dwarfs for comparison. The minimum amd maximum observed $z_o$ values are taken from \cite{Vennes2017, Hardy2023, Hardy2023a, Hollands2023}}. {The red star (J0017+004) corresponding to $z_o=0.4$ leads to near zero asymmetry parameter values; it would be very hard to detect such a magnetic field geometry from asymmetries, and one would have to rely only on the global shifts $\omega_B^{\ell}$. The blue star (J0942+2052), corresponding to an offset of $|z_o|=0.01$ is close to be degenerated with a centered dipole, but the three asymmetry parameters would each lead to a different $\beta_{\ell|m|}$ angle. 
 To understand this in the more general case, we calculate all asymmetry parameters $a_{11}(z_o), a_{21}(z_o), a_{2}(z_o)$ for a fixed offset $z_o$. Then, we calculate for each asymmetry parameter the corresponding inclination angles $\beta_{\ell|m|}$. These three inclination angles for different $a_{\ell|m|}$ might not be the same, and their differences are represented in Fig.~\ref{fig:degeneracies_incl_offset}. When the difference in angles is non-zero the combination of asymmetry parameters is unique to an offset dipole and can be distinguished from an inclined dipole, which is the case for J0942+2052 but not for J0017+0041 which is fully degenerate. This result has to be discussed in perspective with current uncertainties on the measure of asymmetry parameters (see Appendix \ref{sec: a_uncertainty}). For instance, observations from \cite{Li2022} lead to an uncertainty in the inclination angle of $\sim \pm 7°$ \citep[see also][]{Mathis2023}, which is higher than the largest difference in $\beta$ angles measured from asymmetry parameters on Fig.~\ref{fig:degeneracies_incl_offset}. As a result, even though asymmetry parameters are not perfectly identical from the two configurations, it might be complicated to distinguish between a centered dipole and an offset one based on asymmetry parameter values with associated observational uncertainties.}

{The average shift $\omega_{B}^\ell$ depends on the squared averaged magnetic field strength at the H-shell as well, which makes it sensitive to only this particular layer. Strong offset magnetic fields might, therefore, be confused for weaker centered fields when using $\omega_{B}^\ell$ and $a_{\ell|m|}$. This implies that magnetic fields might be underestimated in red giant cores, as a stronger offset dipole can have the same signature on the $\ell=1$ asymmetry parameter than a {weaker} centered dipole (valid also for $\ell=2$). As a result, one should be careful when extracting a magnetic field amplitude from $\omega_{B}^\ell$, which should rather be taken as a minimum amplitude of the large-scale field.} 
 
 %Thus, until the resolution in $a_{\ell, |m|}$ is increased, we cannot differentiate between off-centered and inclined dipoles.

% In between those $z_o$ values, we can draw the same conclusion; even though the asymmetry parameters are not perfectly identical from the two configurations, it might be complicated to distinguish between a weak-centered dipole and a stronger off-centered one based on asymmetry parameter values with associated observational uncertainties.

\begin{table*}[ht]
\centering
\caption{Approximated access to asymmetry parameters depending on the inclination of the star with the line of sight, following mode visibilities as in \cite{Gizon2003}.}
\begin{tabular}{ c c c c c c c c c c c c}
% \hline\hline
 i (°)& \,0\, &  & 10 & & 40 & &70& &80 & &90 \\ [0.5ex] % inserts table %heading
\hline \\
$a_{11}$&\tikzmark{z}{}& \textcolor{gray}{\xmark} &\tikzmark{a}{}& \cmark &\tikzmark{b}{}& \cmark &\tikzmark{c}{}& \cmark &\tikzmark{d}{}& \textcolor{gray}{\xmark}&\tikzmark{e}{}\\
$a_{21}$&& \textcolor{gray}{\xmark} &  &\cmark &  &\textcolor{gray}{\xmark} & & \cmark & & \textcolor{gray}{\xmark}\\
$a_{22}$&& \textcolor{gray}{\xmark}&  & \textcolor{gray}{\xmark}&  & \textcolor{gray}{\xmark}&  & \cmark& & \cmark\\
$a_{22} - a_{21}$&\tikzmark{zz}{}& \textcolor{gray}{\xmark}& \tikzmark{aa}{} & \textcolor{gray}{\xmark} & \tikzmark{bb}{}& \cmark & \tikzmark{cc}{}& \cmark& \tikzmark{dd}{}& \textcolor{gray}{\xmark}&\tikzmark{ee}{}\\ [1ex]
% \hline
\end{tabular}
    \DrawLine{a}{aa}
    \DrawLine{b}{bb}
    \DrawLine{c}{cc}
    \DrawLine{d}{dd}
    \DrawLine{e}{ee}
    \DrawLine{z}{zz}
\label{table:asym}
\end{table*}
\subsection{Accessing asymmetry parameters}

As demonstrated by \cite{Gough1990a} and more recently by \cite{Loi2021}, magnetic fields with a high inclination with respect to the rotation axis generate a second lift of degeneracy of the mixed mode frequencies in the observer frame. The relative amplitude of these additional components of the mixed-mode multiplets compared to the central peaks studied here strongly depends on the magnetic field amplitude. As we place our study in the case of rotation-dominated signatures, these additional multiplet components contain a small fraction of the mode power density \citep{Loi2021}, and therefore have been neglected in our study. As a second result of this approximation, we can neglect the effect of magnetic fields on the mode amplitudes. Therefore, we assume here that the amplitudes in mixed mode multiplets are consistent with the study of \cite{Gizon2003} in the case of pure rotation \citep[further justified by the study of][in the case of weak magnetic fields]{Loi2021}. \\

%Below, we discuss which asymmetry parameters can be accessed for a given line of sight.

As discussed in \cite{Gizon2003}, all components of a given $(\ell, m)$ multiplet are not visible simultaneously, depending on the line of sight of the observation. This makes the detectability of complex magnetic fields challenging, as we cannot access all asymmetry parameters simultaneously. Following Appendix~\ref{app:matrix}, assuming that magnetic fields are at play, the average shift of the $\ell=2$ multiplet $\omega_{B}^{\ell=2}$ can always be estimated when two out of the three $|m|$ components are visible. As a result, for a given inclination angle $i\gtrapprox 10\degree$, at least two components of the $\ell=2$ mixed mode multiplet are visible \citep{Gizon2003}, which result in a systematic estimate for $\omega_{B}^{\ell}$ (i.e. the denominator in Eq.~\ref{eqn: asymmetry_parameter}) when $i\gtrapprox 10\degree$. However, only asymmetry parameters corresponding to the visible components of the multiplet are simultaneously measurable, which depends on $i$. %, even in the case of a combined study of $\ell=1,2$ modes. 
In Table~\ref{table:asym} we report the detectability of the various asymmetry parameters given the inclination of the line of sight $i$ with respect to the rotation axis. Table~\ref{table:asym} shows that we have access to two asymmetry parameters (or of their combination) for most of the inclination angles of the observations ([$a_{11}$ and $a_{21}$] for $10\lessapprox i\lessapprox 40\degree$, [$a_{11}$ and $a_{22}-a_{21}$] for $40\lessapprox i\lessapprox 70\degree$, and [$a_{21}$ and $a_{22}$] for $70\lessapprox i\lessapprox 80\degree$). From Fig.~\ref{fig:3star_degeneracy} we observe that having access to only one out of the two $\ell=2$ asymmetry parameters or to their combination is theoretically enough as $a_{11}$, $a_{21}$, and $a_{22}$ are not independent (see Appendix~\ref{app:matrix}). In a realistic scenario, depending on errors in the measure of the visible asymmetries, two asymmetry parameters might be enough to partially lift the topology degeneracy. The resulting degeneracy depends on the true $(\cR, \beta)$ combination (see Fig.~\ref{fig:3star_degeneracy} in the case of the aligned quadrudipole). We conclude that topologies might be unveiled from $\ell=1,2$ frequencies following our study for stars observed with an inclination angle $10\lessapprox i\lessapprox 80\degree$ when assuming rotation effects dominating magnetic effects. The full investigation of the effect of stronger magnetic fields on the amplitudes and detectability of the different components of the multiplet given the line of sight will the the scope of a follow-up paper.

\section{Conclusion \& Perspectives}
\label{sec:ccl}

{We demonstrate that a combined analysis of $\ell=1$ and $\ell=2$ mixed mode frequency {asymmetries} is key to accessing the quadrupolar component of magnetic fields. When using asymmetry parameters associated with $(\ell=1, |m|=1)$, $(\ell=2, |m|=1)$, and $(\ell=2, |m|=2)$ frequencies, the degeneracy between the signature of aligned and centered dipolar and quadrupolar components of the field can be lifted, allowing to measure their respective strength (depending on the resolution in the data). Aligned quadrupolar fields can therefore be detected from the study of $\ell=2$ oscillation modes. Depending on the inclination of the quadrupole with respect to the dipole, a misalignment might or might not be constrained from $\ell=1$ and $\ell=2$ frequencies.}

{As observed magnetic fields in white dwarfs and main-sequence intermediate stars show offset fields \citep[e.g.][]{Wickramasinghe2000, Hardy2023a}, we also investigated the detectability of such topologies. We demonstrate that strong offset fields can be confused with weaker and centered fields and that we do not currently have a way to distinguish between them. As a result, magnetic field amplitudes estimated from shifts in the frequency pattern should be considered a lower boundary for the true magnetic field amplitude in the radiative zone.}

{Depending on the inclination of the rotation axis of the star with respect to the line of sight, some asymmetry parameters might not be measurable due to low amplitude in $|m|$ components \citep{Gizon2003, Gehan2021}. Our results therefore apply for stars observed with a line of sight $i\in[10,80]\degree$ (which is the case for most observed stars) and for rotation effects dominating magnetic effects, which is the case for magnetic red giants detected so far \citep{Li2022, Deheuvels2022, Li2023}.}
%textcolor{blue}{does the magnetic inclination impact the amplitude of the modes? to investigate}. For low inclination angle ($i\lessapprox 30°$) the \\

{While we derive these magnetoasteroseismology prescriptions for both $\ell=1$ and $\ell=2$ modes, $\ell=2$ mixed oscillation frequencies are extremely complicated to identify in asteroseismic data \citep[e.g.][]{Ahlborn2020}. For this reason, there has not been a dedicated quest for the characterization of quadrupolar mixed modes in the thousands of red giants observed by \textsl{Kepler}, which is lacking in the literature on red giant stars. Considering stars for which rotation and magnetic fields have been measured from $\ell=1$ oscillations, forward modeling of rotating and magnetic mixed-mode patterns including magnetic effects of topologies discussed above for $\ell=2$ modes could be the solution to identify and take advantage of $\ell=2$ frequencies for a better constraint on angular momentum transport inside stars.}

% {Finally; what can we detect without degeneracy?}

% A paragraph about detection in real data. Most likely for young red giant stars, $\nu_{\rm max}<100 \mu$Hz \citep{grosjean2014, deheuvels2020}

% Difficulty to measure $\ell=2$ frequencies, moderate the work by saying that it will take time to be able to detect magnetic effects on qudrupolar mixed mode frequencies.

\begin{acknowledgements}
    The authors thank S. Mathis, L. Barrault, S. Torres, A. Cristea, and K. M. Smith for very useful discussions. This project has received funding from the European Union’s Horizon 2020 research and innovation programme under the Marie Skłodowska-Curie grant agreement No 101034413. 
\end{acknowledgements}

\bibliographystyle{aa} % style aa.bst

\bibliography{references_zotero} % your references Yourfile.bib

\appendix
\allowdisplaybreaks
\section{$B_r$ expression for an inclined quadrudipole} \label{sec:Br_dip_quad}
We briefly outline the mathematical steps to derive the expression of $B_r$ for a general quadrudipole, where the dipolar and quadrupolar axes are not necessarily aligned. The schematic representation of a general quadrudipole is shown in Fig.~\ref{fig:field_config_panel}. The dipole is inclined at an angle $\beta_d$ to the rotation axis and the quadrupole is inclined at an angle $\beta_q$ to the rotation axis. To go about constructing this, let's start at the simpler case where the symmetry axis of the dipole and quadrupole are aligned with the rotation axis. The expression for the radial field component in this case will look like
\begin{equation}
    B_r(r, \theta, \phi) = B_0 \, b_r(r) \left[Y_{10}(\theta, \varphi) + \cR \, Y_{20}(\theta, \varphi)\right],
\end{equation}
where $\cR$ is the ratio of the strength of the quadrupole to the dipole. Now, we want to incline the dipolar component by an angle $\beta_d$ with respect to the rotation axis. For this, let's first choose a frame $(\Tilde{\theta}_d, \Tilde{\varphi}_d)$ where the dipole is axisymmetric and then find the coordinate transform between $(\Tilde{\theta}_d, \Tilde{\varphi}_d)$ and $(\theta, \varphi)$. Thanks to Wigner $d$-matrices \citep[][]{Varshalovich1988}, we can use the following relation to go from the axisymmetric frame to a non-axisymmetric frame rotated by $\beta$
\begin{equation} \label{eq: sph_Wid_d}
    Y_{\ell, m}(\theta, \varphi) = d^{(\ell)}_{m,0}(\beta_d) \, Y_{\ell,0}(\Tilde{\theta}_d, \Tilde{\varphi}_d) \, .
\end{equation}
Now, using the fact that Wigner $d$-matrices are unitary, we get the expression
\begin{equation}
    Y_{\ell, 0}(\Tilde{\theta}_d, \Tilde{\varphi}_d) = \sum_{m=-\ell}^{\ell} d^{(\ell)}_{0,m}(\beta_d) \, Y_{\ell,m}(\theta, \varphi) \, .
\end{equation}
So, for a dipole inclined at $\beta_d$ from the rotation axis and a quadrupole aligned with the rotation axis, the expression for the radial component of the magnetic field reads
\begin{align}
    B_r(r, \theta, \phi) &= B_0 \, b_r(r) \left[Y_{10}(\Tilde{\theta}_d, \Tilde{\varphi}_d) + \cR \, Y_{20}(\theta, \varphi)\right] \, , \nonumber \\
    &= B_0 \, b_r(r) \left[\sum_{m=-1}^{1} d^{(1)}_{0,m}(\beta_d) \, Y_{1,m}(\theta, \varphi) + \cR \, Y_{20}(\theta, \varphi)\right] \, .
\end{align}
Similarly, the transformation between a coordinate that is rotated by $\beta_q$ (where the quadrupole is axisymmetric) and the coordinate of axisymmetric rotation is given by
\begin{equation}
    Y_{\ell, 0}(\Tilde{\theta}_q, \Tilde{\varphi}_q) = \sum_{m=-\ell}^{\ell} d^{(\ell)}_{0,m}(\beta_q) \, Y_{\ell,m}(\theta, \varphi) \, .
\end{equation}
Therefore, the total $B_r (r, \theta, \varphi)$ where the dipole and quadrupole are tilted by $\beta_d$ and $\beta_q$ with respect to the rotation axis is given by
\begin{align} \label{eqn: derive_general_quadrudipole}
    B_r(r,\theta,\varphi) = B_0 \, b_r(r) \Bigg(\sum_{m=-1}^{1} & d^{(1)}_{0,m}(\beta_d) \, Y_{1m}(\theta, \varphi) \nonumber \\
    +& \cR \sum_{m=-2}^{2} d^{(2)}_{0,m}(\beta_q) \, Y_{2m}(\theta, \varphi)\Bigg),
\end{align}
On plugging in the elements of the Wigner $d$-matrices, in Eq.~\ref{eqn: derive_general_quadrudipole}, we get the expression in Eq.~\ref{eqn: quadrudipole_Br_expanded}.

% \begin{table*}[b]
% \centering
% \caption{Approximated access to oscillation frequencies $\omega_{\ell, m}$ depending on the inclination of the star with the line of sight, following mode visibilities as in \cite{Gizon2003}.}
% \begin{tabular}{ c c c c c c c c c c c c}
% % \hline\hline
%  i (°)& \,0\, &  & 10 & & 40 & &70& &80 & &90 \\ [0.5ex] % inserts table %heading
% \hline \\
% $\omega_{1,0}$&\tikzmark{z}{}& {\cmark} &\tikzmark{a}{}& \cmark &\tikzmark{b}{}& \cmark &\tikzmark{c}{}& \cmark &\tikzmark{d}{}& \textcolor{gray}{\xmark}&\tikzmark{e}{}\\
% $\omega_{1,1}$&& \textcolor{gray}{\xmark} &  &\cmark &  &{\cmark} & & \cmark & & {\cmark}\\
% $\omega_{2,0}$&& {\cmark}&  & {\cmark}&  & \textcolor{gray}{\xmark}&  & \textcolor{gray}{\xmark}& & \textcolor{gray}{\xmark}\\
% $\omega_{2,1}$&\tikzmark{zz}{}& \textcolor{gray}{\xmark}& \tikzmark{aa}{} & {\cmark} & \tikzmark{bb}{}& \cmark & \tikzmark{cc}{}& \cmark& \tikzmark{dd}{}& \textcolor{gray}{\xmark}&\tikzmark{ee}{}\\
% $\omega_{2,2}$&\tikzmark{zz}{}& \textcolor{gray}{\xmark}& \tikzmark{aa}{} & \textcolor{gray}{\xmark} & \tikzmark{bb}{}& \cmark & \tikzmark{cc}{}& \cmark& \tikzmark{dd}{}& {\cmark}&\tikzmark{ee}{}\\ [1ex]
% % \hline
% \end{tabular}
%     \DrawLine{a}{aa}
%     \DrawLine{b}{bb}
%     \DrawLine{c}{cc}
%     \DrawLine{d}{dd}
%     \DrawLine{e}{ee}
%     \DrawLine{z}{zz}
% \label{table:detect_modes}
% \end{table*}

\section{Magnetic inversion kernels} \label{sec: Brsq_kerns}
Section~\ref{sec: magsplitpy} outlines the numerical approach we adopt to calculate the magnetic coupling matrix. A key component in the numerical evaluation of magnetically perturbed stellar eigenstates is its Lorentz-stress sensitivity kernels \citep[originally laid out in][]{Das2020}. These Lorentz-stress kernel components ${}_{k'k}\cB_{st}^{\mu\nu}$ in Eq.~\ref{eqn:lamda_decomp}, are defined as
 \begin{equation}
    {}_{k'k}\cB_{st}^{\mu\nu}  = 4\pi (-1)^{m'}\gamma_{\ell'} \gamma_{s} \gamma_{\ell}\wigred{-m'}{t}{m}  \, {}_{k'k}\cG_{s}^{\mu\nu} \, , \label{eqn: m_ind_kern}
\end{equation}
where, $\gamma_{\ell} = \sqrt{2\ell+1 / 4\pi}$ and $\wigred{-m'}{t}{m}$ are the Wigner 3-$j$ symbols obtained from triple product of complex spherical harmonics \citep[see Appendix~C of][]{DT98}. The $m$ independent part of the kernels ${}_{k'k}\cG_{s}^{\mu\nu}$ are functions of the stellar eigenfunctions (hence the structure of the background stellar model)

\begin{eqnarray}
\cG_{s}^{--} = &&\frac{-1}{2r^2}  \Bigg[\wigred{2}{-2}{0} \chi_1^{--}(k,k') +\wigred{0}{-2}{2} \chi_1^{--}(k',k) \nonumber\\ && +~\wigred{1}{-2}{1} \big\{ \chi_2^{--}(k,k') + \chi_2^{--}(k',k) \big\} \nonumber\\ && +~\wigred{3}{-2}{-1} \chi_3^{--}(k,k') + \wigred{-1}{-2}{3} \chi_3^{--}(k',k)\Bigg],\label{eq:kern_mm} \\
\cG_{s}^{0-} = &&\frac{1}{4r^2} \Bigg[\wigred{1}{-1}{0} \chi_1^{0-}(k,k') + \wigred{0}{-1}{1}\chi_1^{0-}(k',k) \nonumber\\ &&+~\wigred{-1}{-1}{2} \chi_2 ^{0-}(k,k') +~\wigred{2}{-1}{-1}\chi_2^{0-}(k',k)\Bigg],\label{eq:kern_0m} \\
\cG_{s}^{00} = &&\frac{1}{2r^2} (1+p) \bigg\{\tfrac{1}{2}\,\wigred{0}{0}{0} \left[\chi_1^{00}(k,k') + \chi_1^{00}(k',k) \right] \nonumber\\ &&+ \wigred{-1}{0}{1} \left[\chi_2^{00}(k,k') + \chi_2^{00}(k',k) \right] \bigg\}, \nonumber\label{eq:kern_00} \\
\cG_{s}^{+-} = &&\frac{1}{4r^2} (1+p) \bigg\{\tfrac{1}{2}\,\wigred{0}{0}{0} \left[ \chi_1^{+-}(k,k') + \chi_1^{+-}(k',k)\right] \nonumber \\ &&+~\wigred{-2}{0}{2} \left[ \chi_2^{+-}(k,k') + \chi_2^{+-}(k',k) \right] \nonumber\\ && +~\wigred{-1}{0}{1}  \left[\chi_3^{+-}(k,k') +  \chi_3^{+-}(k',k) \right] \bigg\}. \label{eq:kern_pm}
\end{eqnarray}
where $p = (-1)^{\ell+\ell'+s}$ and 

\begin{align}
    \chi_{1}^{--}(k) &= \Omega_{0\ell}\Omega_{2\ell} \left[ V_k(3U_k-2\Omega_{2\ell}^{2} V_k +3r\dot{U_k})-rU_k\dot{V_k}\right],
    \\
    \chi_{2}^{--}(k) &= \Omega_{0\ell}^2 
    \left[
        3U_kV_k + (\Omega_{2\ell}^2- 2\Omega_{0\ell}^{2})V_k^{2} + rV_k\dot{U_k} - rU_k\dot{V_k} - U_k^{2}
    \right],
    \\
    \chi_{3}^{--}(k) &= \Omega_{0\ell}^2 \Omega_{2\ell} \Omega_{3\ell} V_k^2,
    \\
    \chi_{1}^{0-}(k)
    &=
    \Omega_{0\ell}
    \Big[
        4\Omega_{0\ell}^2 V_k^2 - 4r\Omega_{0\ell}^2 V_k\dot{V_k} + 2r^2 \dot{U_k}\dot{V_k} + r^2 V_k \ddot{U_k} \nonumber \\
        & \hspace{1cm} + U_k\lbrace 8U_k-6(\Omega_{0\ell}^2+1)V_k+r(4\dot{V_k}-r\ddot{V_k}) \rbrace \Big], \\
    \chi_{2}^{0-}(k)
    &= \Omega_{0\ell}^2\Omega_{2\ell} \left[ U_kV_k + V_k(U_k-4V_k+3r\dot{V_k}) + rV_k\dot{V_k} \right],
    \\
    \chi_{1}^{00}(k) &= 2
    \Big[
        -2rU_k\dot{U_k} + \Omega_{0\ell}^{2}r(V_k\dot{U_k} + U_k \dot{V_k}) \nonumber \\
        & \hspace{1cm} - 5\Omega_{0\ell}^2 V_kU_k + 2\Omega_{0\ell}^4 V_k^{2} + 3U_k^{2}
    \Big],
    \\
    \chi_{2}^{00}(k) &=
    -\Omega_{0\ell}^2 
    \Big[
        - U_kV_k + V_k^2 + r(V_k\dot{U_k}+U_k\dot{V_k}) \\
        &- 2rV_k\dot{V_k} + r^{2}\dot{V_k}^2
    \Big],
    \\
    \chi_{1}^{+-}(k) &= 2 
    \Big[
        -2r\dot{U_k}U_k + \Omega_{0\ell}^2 r( \dot{U_k}V_k + U_k\dot{V_k} ) - r^2 \dot{U_k}^2 \nonumber \\
        & \hspace{1cm} - \Omega_{0\ell}^2U_k V_k + U_k^2
    \Big],
    \\
    \chi_{2}^{+-}(k) &= -2\Omega_{0\ell}^2\Omega_{2\ell}^2 V_k^2,
    \\
    \chi_{3}^{+-}(k) &= \Omega_{0\ell}^2 
    \left[
        r(U_k\dot{V_k}-V_k\dot{U_k}) - U_k V_k + U_k^2
    \right],
\end{align}
and $\Omega_{N\ell} = \sqrt{\frac{1}{2}(\ell + N)(\ell - N + 1)}$. From visual inspection, the coupling matrix $M_{k'k}$ (built on the kernels) is hermitian which ensures real eigenfrequencies. Further, the $(1+p)$ factor in Eq.~\ref{eq:kern_00} \& \ref{eq:kern_pm} implies that for self-coupling of multiplets, the frequency splittings are sensitive to only even $s$ components of the Lorentz-stress components $h_{st}^{00}$ and $h_{st}^{+-}$ arising from $B_r^2$ and $(B_{\theta}^2 + B_{\varphi}^2)$, respectively. 

 \section{Elements and trace of magnetic coupling matrix for $\ell=2$}
\label{app:matrix}
\subsection{Elements of the magnetic coupling matrix for $\ell=2$ modes}
We followed the same steps from Eq.~30 of \cite{Li2022} to find the elements of the magnetic coupling matrix $\mathbf{M}_{\ell}$ in the case of $\ell=2$ mixed modes:

\begin{eqnarray*}
    M_2^{2,2} = M_2^{-2,-2} &=& \frac{1}{2 \, \mu_0 \, \omega \, I} \frac{15}{4} \int_{r_i}^{r_o} \left[\frac{\partial(r \xi_h)}{\partial r}\right]^2 \int_{0}^{\pi} B_r^2 \\
    &&\, \{ \sin^2{\theta} \, \cos^2{\theta}+ \sin^2{\theta} \} r^2 \, \sin{\theta} \, dr \, d\theta \\
    M_2^{1,1} = M_2^{-1,-1} &=& \frac{1}{2 \, \mu_0 \, \omega \, I} \frac{15}{4} \int_{r_i}^{r_o} \left[\frac{\partial(r \xi_h)}{\partial r}\right]^2 \int_{0}^{\pi} B_r^2 \\
    &&\, \{ \cos^4{\theta} + \sin^4{\theta} \, - 2 \, \cos^2{\theta} \, \sin^2{\theta} + \cos^2{\theta} \} \\
    &&r^2 \, \sin{\theta} \, dr \, d\theta \\
    M_2^{0,0} &=& \frac{1}{2 \, \mu_0 \, \omega \, I} \frac{45}{2} \int_{r_i}^{r_o} \left[\frac{\partial(r \xi_h)}{\partial r}\right]^2 \int_{0}^{\pi} B_r^2 \\
    &&\, \sin^2{\theta} \, \cos^2{\theta} \, r^2 \, \sin{\theta} \, dr \, d\theta 
\end{eqnarray*}

\subsection{Obtaining net magnetic shift in ${\ell=2}$ from net magnetic shift of $\ell=1$}
Focusing on the angular part of the integrand other than the dependence from $B_r^2$, we see that
\begin{eqnarray}
    M_2^{2,2} &\propto& \sin^2{\theta} \, \cos^2{\theta} + \sin^2{\theta} \, \\
    M_2^{1,1} &\propto& 1 - 4 \, \cos^2{\theta} \, \sin^2{\theta} + \cos^2{\theta} \\
    %\cos^4{\theta} + \sin^4{\theta} \, - 2 \, \cos^2{\theta} \, \sin^2{\theta} + \cos^2{\theta} \, \nonumber \\
    % &=& 1 - 4 \, \cos^2{\theta} \, \sin^2{\theta} + \cos^2{\theta} \\
    M_2^{0,0} &\propto& 6 \, \sin^2{\theta} \, \cos^2{\theta}
\end{eqnarray}
Using simple trigonometric identities, it is easy to see that this angular part in the trace of $\mathbf{M}_{2}$ evaluates to $\mathrm{Tr}(\mathbf{M}_{2}) = 2 \, M_2^{2,2} + 2\, M_2^{1,1} + M_2^{0,0} \propto 4$. The complete expression of the net magnetic shift then becomes
\begin{equation} \label{eqn: Trace_M}
    \mathrm{Tr}(\mathbf{M}_{2}) = \frac{15}{2 \, \mu_0 \, \omega \, I} \int_{r_i}^{r_o} \left[\frac{\partial(r \xi_h)}{\partial r}\right]^2 \int_{0}^{\pi} B_r^2 \, \sin{\theta} \, dr \, d\theta \, .
\end{equation}
% Since $\ell=1$ mode splittings are more detectable than $\ell=2$ modes in RGB, it is fair to assume that for most cases, we will have access to magnetic splittings and shifts of dipolar modes. 

\noindent All $\ell=1$ mode frequencies are detectable in the range $i\in[10,80]\degree$ \citep{Gizon2003}. \cite{Li2022} used the net shift in $\ell=1$ modes $\omega_B^{\ell=1}$ to obtain horizontal average of the squared average magnetic field
\begin{equation} \label{eqn: avg_Br_sq}
    \overline{B_r^2} \sim \int_0^{\pi} B_r^2 \, \sin{\theta} \, d\theta \, .
\end{equation}
Therefore, from an independent analysis of the $\ell=1$ modes of the same star with $i\in[10,80]\degree$, in the first step we can estimate $\overline{B_r^2}$ from the $\ell=1$ modes which we can then use to calculate $\mathrm{Tr}(\mathbf{M}_2)$ even if we only have explicit access to two of the $\omega_{2 |m|}$ due to mode visibility induced by a relative inclination between the rotation and magnetic axes. Using Eq.~\ref{eqn: asymmetry_parameter}, Eq.~\ref{eqn: Trace_M}, and Eq.~\ref{eqn: avg_Br_sq}, we get the following relation between the three asymmetry parameters
\begin{equation}
    a_{22} + a_{21} + \frac{5 \, M_{2}^{0,0}}{\mathcal{T} \, \overline{B_r^2}(a_{11})} = 1 \, .
\end{equation}
where the radial dependence of the mode sensitivity around the H-shell is captured in
\begin{equation} \label{eqn: prefactor}
    \mathcal{T} =  \frac{15}{2 \, \mu_0 \, \omega \, I} \int_{r_i}^{r_o} \left[\frac{\partial(r \xi_h)}{\partial r}\right]^2 \, dr \, .
\end{equation}

% \textcolor{red}{As a result of this relation between the elements of the trace of the coupling matrix, CAN WE SAY THAT IF WE HAVE TWO OUT OF THE THREE asymmetry parameters we get the third one?}

\section{Numerical calculation for Case (b)}
\label{sec:numerical_misaligned}
% \textbf{To be moved to Appendix?}
{Similar to section \ref{sec:numerical_params_aligned} we calculate the asymmetry parameters for a magnetic field configuration of the case (b) in Fig. \ref{fig:field_config_panel}. The quadrupolar field in this case is chosen to be inclined by $90\degree$ compared to the dipolar field. The resulting comparison between the numerical results and the analytical expressions can be found in Fig. \ref{fig:magsplitpy_params_misaligned}. Discrepancies between the numerical and the analytical method are once again negligible, as in Case (a).}\\

\begin{figure*}
    \centering
    \includegraphics[width=0.94\textwidth]{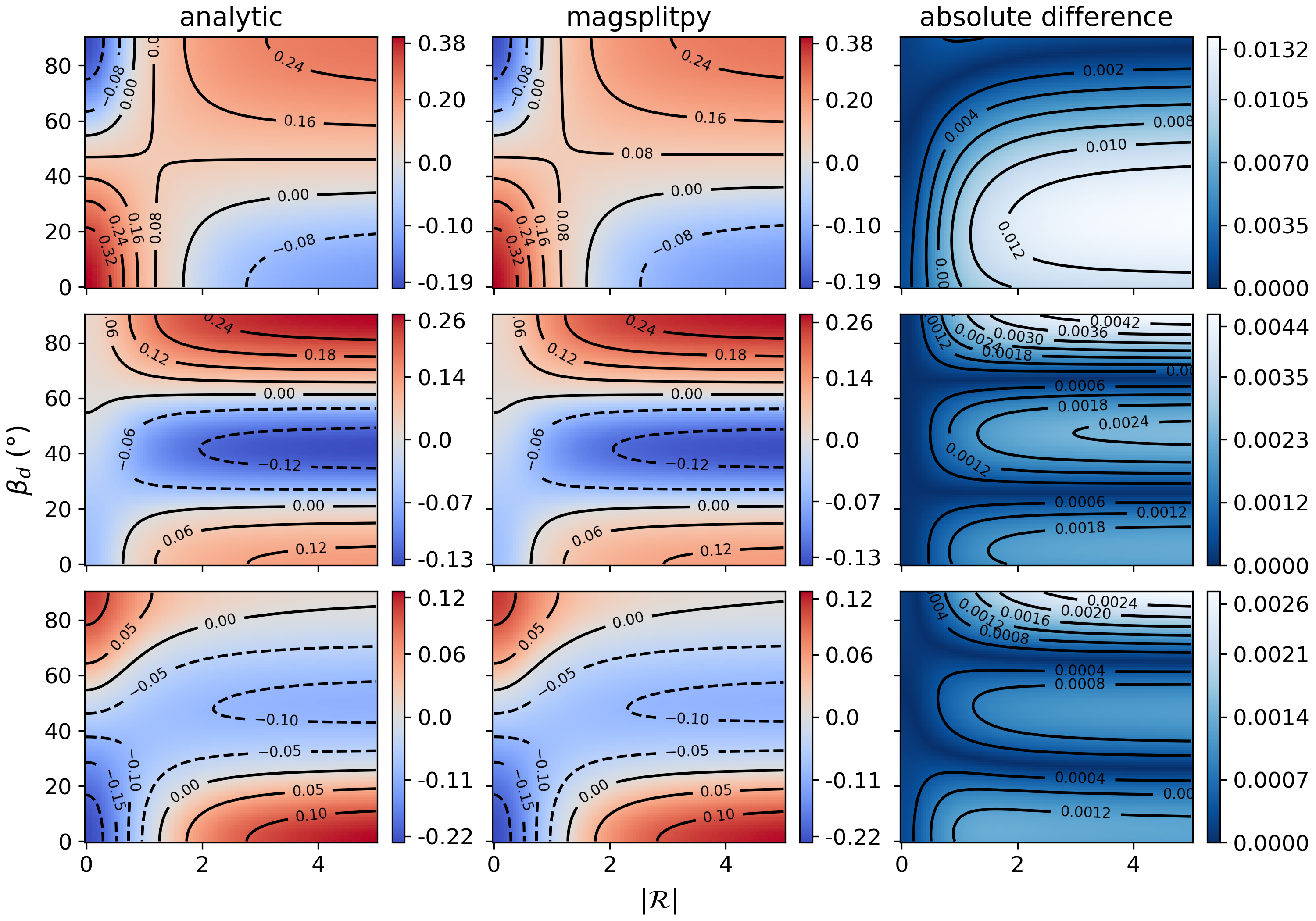}
    \caption{Same as Figure~\ref{fig:magsplitpy_params_aligned} in the case (b) of a misaligned quadrudipole in the case where $\beta_q-\beta_d=90\degree$.}
    \label{fig:magsplitpy_params_misaligned}
\end{figure*}

{Fig.~\ref{fig:3d_projection_2} is an extension of Fig.~\ref{fig:3d_projection_degeneracy_1}. We use the asymmetry parameter values corresponding to a Case (a) field with $(\beta\in [0:90]\degree, |\cR|\in [0.5, 1.1, 1.8])$ and show the also possible ($|\cR|$, $\beta_d$, $\beta_q$) parameters corresponding to a Case (b) field leading to the same asymmetry values. Along the y-axis we vary the initial angle on the aligned quadrudipole (Case (a) $\beta=\beta_d=\beta_q$. For initial $\beta$ of 0° and 90° the values of $|\cR|$, $\beta_d$, $\beta_q$ are unique. If a chosen star has an inclination angle $\beta$ between 35° and 65°, the dipole angle is fully degenerate. The possible values $|\cR|$, $\beta_d$, $\beta_q$ can take for the three stars in Fig.~\ref{fig:3d_projection_degeneracy_1} are shown as the horizontal red lines, the red symbols indicating the case where $\beta_d=\beta_q=\beta$ corresponding to the aligned field as in Fig.~\ref{fig:3star_degeneracy}. Star 1 can either be i) an aligned quadrudipole with $\beta=25\degree$ and $|\cR|=0.5$ or ii) a misaligned quadrudipole with $\beta_d\lessapprox 27\degree$, $\beta_q\lessapprox 29\degree\, \mathrm{or}\, \gtrapprox 77\degree$ and $|\cR|\in[0.2:1.2]$. Star 2 is very well constrained as an aligned quadrudipole, with very small $\beta$. For Star 3 we cannot constrain the dipole angle, but we know the quadrupole with an angle between $45\degree$ and $58\degree$ must dominate the field strength.} 

\begin{figure*}
    \centering
    \includegraphics[width=0.94\textwidth]{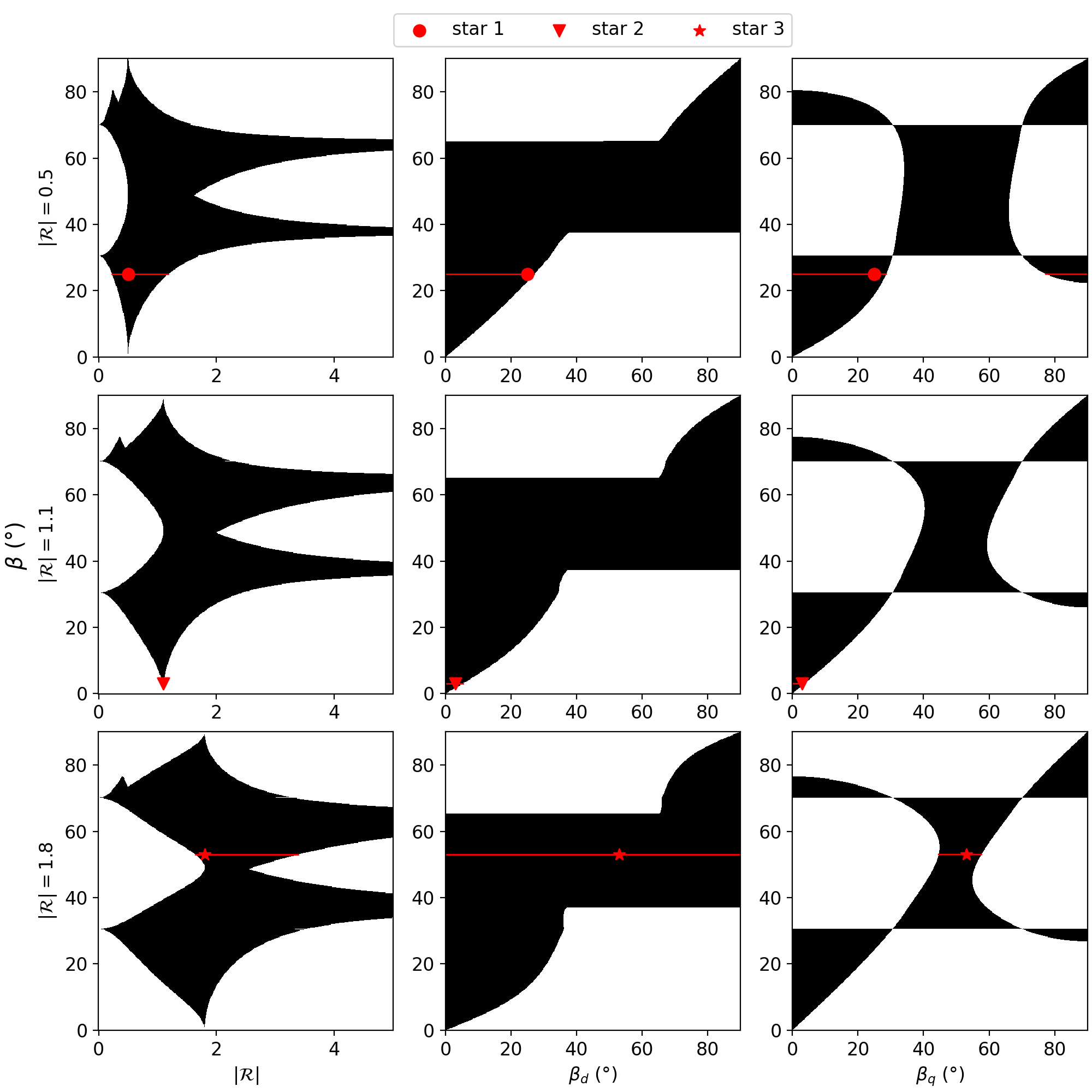}
    \caption{Degeneracies in the asymmetry parameters for the three dipole to quadrupole strength ratios corresponding to Star 1 to 3, as a function of the original $\beta$ angle and $|\cR|$ ratio. }
    \label{fig:3d_projection_2}
\end{figure*}

\section{Observational uncertainty on asymmetry parameters} \label{sec: a_uncertainty}
From Eq.~\ref{eqn: asym_param_def} we know that $\omega_{\ell, -m} + \omega_{\ell, m} - 2 \omega_{\ell, 0} = (2\ell+1) \zeta \, a_{\ell |m|}  \, \omega_{B}^\ell$, where our definition follows the convention of \cite{Li2022}, with $\sum_{m=-\ell}^{\ell} \delta \omega_{\ell, m} = (2\ell+1) \zeta \,\omega_{B}^\ell= (2\ell+1) \delta\omega_{B}^\ell$. Consequently, we can write the asymmetry parameter in the form
\begin{equation}
    a_{\ell |m|} = \frac{\delta \omega_{\ell, -m} + \delta \omega_{\ell, m} - 2 \delta \omega_{\ell, 0}}{\sum_{m=-\ell}^{\ell} \delta \omega_{\ell, m}} =\frac{N}{D}\, .
\end{equation}
Here, we have used the fact that the total perturbed frequency $\omega_{\ell, m}$ is a sum of the unperturbed degenerate frequency $\omega_{\ell}^0$ and the splitting $\delta \omega_{\ell, m}$. In the above expression, the numerator $N$ and the denominator $D$ represent measurable quantities from the asteroseismic power spectra. We then use the error propagation 
\begin{equation} \label{eqn: err_prop_ratio}
    \frac{\delta a_{\ell|m|}}{a_{\ell|m|}} = \sqrt{\left(\frac{\delta N}{N}\right)^2 + \left(\frac{\delta D}{D}\right)^2} \, .
\end{equation}
To estimate the errors in N and D, we use the error propagation for summations. This gives us $\delta N = \sqrt{6} \, \sigma_{\omega}$ and $\delta D = \sqrt{3} \, \sigma_{\omega}$ where $\sigma_{\omega}$ is the uncertainty in measuring mode frequencies from data. Plugging these into Eq.~\ref{eqn: err_prop_ratio}
\begin{equation} \label{eqn: delta_a_exp1}
    \delta a_{\ell|m|} = \frac{\sigma_{\omega}}{\sqrt{3} \, \delta \omega_{B}^\ell} \sqrt{2 + \left(\frac{\delta \omega_{\ell, -m} + \delta \omega_{\ell, m} - 2 \delta \omega_{\ell, 0}}{\sum_{m=-\ell}^{\ell} \delta \omega_{\ell, m}}\right)^2}
\end{equation}
Since asymmetry parameters are of the order $\mathcal{O}(10^{-1})$ --- for instance $a_{11} \in [-0.2, 0.4]$ in \cite{Mathis2023} for a dipolar field, we can approximate Eq.~\ref{eqn: delta_a_exp1} to
\begin{equation}
    \delta a_{\ell|m|} \sim \sqrt{\frac{2}{3}}\frac{\sigma_{\omega}}{\delta \omega_{B}^\ell} = \frac{\sqrt{6} \, \delta f}{\delta \omega_{B}^\ell} \, ,
\end{equation}
where we assume the minimum reliable frequency resolution $\delta$ is 3 times the satellite data resolution $\sigma_{\omega}$. For 4-year \textit{Kepler} data, $\delta f \equiv 7.2$nHz and for a typical red giant star \citep[as in][]{Li2022} since the average magnetic shift in gravity modes $\omega_{B}^\ell \sim 200$nHz and $\zeta\approx 1$ for gravity-dominated modes, the above equation gives us $\delta a_{\ell|m|} \sim 0.08$.

\end{document}